\documentclass[twocolumn,showpacs,preprintnumbers,amsmath,amssymb,superscriptaddress]{revtex4}

\usepackage{graphicx,color}
\usepackage{dcolumn}
\usepackage{bm}

\begin{document}

\preprint{APS/123-QED}

\title{Hadronic resonance production in $d$+Au collisions at $\sqrt{s_{_{NN}}}$ = 200 GeV at RHIC}

\date{\today}


\affiliation{Argonne National Laboratory, Argonne, Illinois 60439, USA}
\affiliation{University of Birmingham, Birmingham, United Kingdom}
\affiliation{Brookhaven National Laboratory, Upton, New York 11973, USA}
\affiliation{University of California, Berkeley, California 94720, USA}
\affiliation{University of California, Davis, California 95616, USA}
\affiliation{University of California, Los Angeles, California 90095, USA}
\affiliation{Universidade Estadual de Campinas, Sao Paulo, Brazil}
\affiliation{Carnegie Mellon University, Pittsburgh, Pennsylvania 15213, USA}
\affiliation{University of Illinois at Chicago, Chicago, Illinois 60607, USA}
\affiliation{Creighton University, Omaha, Nebraska 68178, USA}
\affiliation{Nuclear Physics Institute AS CR, 250 68 \v{R}e\v{z}/Prague, Czech Republic}
\affiliation{Laboratory for High Energy (JINR), Dubna, Russia}
\affiliation{Particle Physics Laboratory (JINR), Dubna, Russia}
\affiliation{Institute of Physics, Bhubaneswar 751005, India}
\affiliation{Indian Institute of Technology, Mumbai, India}
\affiliation{Indiana University, Bloomington, Indiana 47408, USA}
\affiliation{Institut de Recherches Subatomiques, Strasbourg, France}
\affiliation{University of Jammu, Jammu 180001, India}
\affiliation{Kent State University, Kent, Ohio 44242, USA}
\affiliation{University of Kentucky, Lexington, Kentucky, 40506-0055, USA}
\affiliation{Institute of Modern Physics, Lanzhou, China}
\affiliation{Lawrence Berkeley National Laboratory, Berkeley, California 94720, USA}
\affiliation{Massachusetts Institute of Technology, Cambridge, MA 02139-4307, USA}
\affiliation{Max-Planck-Institut f\"ur Physik, Munich, Germany}
\affiliation{Michigan State University, East Lansing, Michigan 48824, USA}
\affiliation{Moscow Engineering Physics Institute, Moscow Russia}
\affiliation{City College of New York, New York City, New York 10031, USA}
\affiliation{NIKHEF and Utrecht University, Amsterdam, The Netherlands}
\affiliation{Ohio State University, Columbus, Ohio 43210, USA}
\affiliation{Panjab University, Chandigarh 160014, India}
\affiliation{Pennsylvania State University, University Park, Pennsylvania 16802, USA}
\affiliation{Institute of High Energy Physics, Protvino, Russia}
\affiliation{Purdue University, West Lafayette, Indiana 47907, USA}
\affiliation{Pusan National University, Pusan, Republic of Korea}
\affiliation{University of Rajasthan, Jaipur 302004, India}
\affiliation{Rice University, Houston, Texas 77251, USA}
\affiliation{Universidade de Sao Paulo, Sao Paulo, Brazil}
\affiliation{University of Science \& Technology of China, Hefei 230026, China}
\affiliation{Shanghai Institute of Applied Physics, Shanghai 201800, China}
\affiliation{SUBATECH, Nantes, France}
\affiliation{Texas A\&M University, College Station, Texas 77843, USA}
\affiliation{University of Texas, Austin, Texas 78712, USA}
\affiliation{Tsinghua University, Beijing 100084, China}
\affiliation{United States Naval Academy, Annapolis, MD 21402, USA}
\affiliation{Valparaiso University, Valparaiso, Indiana 46383, USA}
\affiliation{Variable Energy Cyclotron Centre, Kolkata 700064, India}
\affiliation{Warsaw University of Technology, Warsaw, Poland}
\affiliation{University of Washington, Seattle, Washington 98195, USA}
\affiliation{Wayne State University, Detroit, Michigan 48201, USA}
\affiliation{Institute of Particle Physics, CCNU (HZNU), Wuhan 430079, China}
\affiliation{Yale University, New Haven, Connecticut 06520, USA}
\affiliation{University of Zagreb, Zagreb, HR-10002, Croatia}

\author{B.~I.~Abelev}\affiliation{University of Illinois at Chicago, Chicago, Illinois 60607, USA}
\author{M.~M.~Aggarwal}\affiliation{Panjab University, Chandigarh 160014, India}
\author{Z.~Ahammed}\affiliation{Variable Energy Cyclotron Centre, Kolkata 700064, India}
\author{B.~D.~Anderson}\affiliation{Kent State University, Kent, Ohio 44242, USA}
\author{D.~Arkhipkin}\affiliation{Particle Physics Laboratory (JINR), Dubna, Russia}
\author{G.~S.~Averichev}\affiliation{Laboratory for High Energy (JINR), Dubna, Russia}
\author{Y.~Bai}\affiliation{NIKHEF and Utrecht University, Amsterdam, The Netherlands}
\author{J.~Balewski}\affiliation{Massachusetts Institute of Technology, Cambridge, MA 02139-4307, USA}
\author{O.~Barannikova}\affiliation{University of Illinois at Chicago, Chicago, Illinois 60607, USA}
\author{L.~S.~Barnby}\affiliation{University of Birmingham, Birmingham, United Kingdom}
% ...

\collaboration{STAR Collaboration}\noaffiliation

\begin{abstract}
We present the first measurements of the $\rho(770)^0$,
$K^*$(892), $\Delta$(1232)$^{++}$, $\Sigma$(1385), and
$\Lambda$(1520) resonances in $d$+Au collisions at
$\sqrt{s_{_{NN}}}$ = 200 GeV, reconstructed via their hadronic
decay channels using the STAR detector at RHIC. The masses and
widths of these resonances are studied as a function of transverse
momentum ($p_T$). We observe that the resonance spectra follow a
generalized scaling law with the transverse mass ($m_T$). The
$\langle p_T \rangle$ of resonances in minimum bias collisions is
compared to the $\langle p_T \rangle$ of $\pi$, $K$, and
$\overline{p}$. The $\rho^0/\pi^-$, $K^*/K^-$, $\Delta^{++}/p$,
$\Sigma(1385)/\Lambda$, and $\Lambda(1520)/\Lambda$ ratios in
$d$+Au collisions are compared to the measurements in minimum bias
$p+p$ interactions, where we observe that both measurements are
comparable. The nuclear modification factors ($R_{dAu}$) of the
$\rho^0$, $K^*$, and $\Sigma^*$ scale with the number of binary
collisions ($N_{bin}$) for $p_T >$ 1.2 GeV/$c$.
\end{abstract}


\maketitle

\section{Introduction}

Quantum chromodynamics (QCD) predicts that hadronic matter at high
temperatures and/or high densities undergoes a phase transition to
a system of deconfined partonic matter, the Quark Gluon Plasma
(QGP) \cite{mil}. Matter under such extreme conditions can be
studied in the laboratory by colliding nuclei at very high
energies. The Relativistic Heavy-Ion Collider (RHIC) at Brookhaven
National Laboratory has provided a variety of collision systems at
different beam energies, including collisions of Au+Au, $d$+Au and
$p+p$ at $\sqrt{s_{_{NN}}}$ = 200 GeV.

Resonances are strongly decaying particles with lifetimes $\times$ $c$
that are
of the order of the size of the hot and dense medium produced in
heavy-ion collisions. As such, the measurement of resonances in
Au+Au collisions compared to $p+p$ collisions at
$\sqrt{s_{_{NN}}}$ = 200 GeV has provided detailed information
about the interaction dynamics in relativistic heavy-ion
collisions \cite{kstar130,rho,kstar200,sevil,sevil1,baryon}, where
hadronic lifetimes and interaction cross-sections affect the
resonance yields
\cite{kstar130,urqmd1,urqmd2,raf1,raf2,raf3,mar02}.

The in-medium effects related to the high density and/or high
temperature of the medium can modify the properties of short-lived
resonances, such as their masses, widths, and even their spectral
shapes \cite{brown,rapp4,shuryak}. Thus, resonances provide a
unique tool for studying various properties of interaction
dynamics in relativistic heavy-ion collisions
\cite{bielich,rapp3}. A good understanding of resonance production
in the reference systems $p+p$ and $d$+Au is useful in
understanding resonance production in Au+Au collisions.
Comparisons between $p+p$ and Au+Au for the $\rho^0$ and $K^{*0}$
mesons have been made elsewhere \cite{rho,kstar200}, where it was
observed that there were modifications of the resonance properties
(mass and width) in both systems with respect to values in the
vacuum in the absence of any medium effects. The measurement
of masses and widths of resonances in $d$+Au collisions add
further information to the existing measurements.

In addition, the regeneration of resonances and the re-scattering
of their daughters are two competing effects that affect the
interpretation of resonance production. Resonances that decay
before kinetic freeze-out (vanishing elastic collisions) may not
be reconstructed due to the re-scattering of the daughter
particles. In this case, the resonance survival probability is
relevant and depends on the time between chemical and kinetic
freeze-outs, the source size, and the resonance transverse
momentum ($p_T$). On the other hand, after chemical freeze-out
(vanishing inelastic collisions), elastic interactions may
increase the resonance population compensating for the ones that
decay before kinetic freeze-out. The case of resonance
regeneration depends on the hadronic cross-section of their
daughters. Thus, the study of resonances can provide an
independent probe of the time evolution of the source from
chemical to kinetic freeze-out
and detailed information on hadronic interaction at later stages.
This has been measured in Au+Au and compared to $p+p$ for the
$K^*$, $\Sigma^*$, and $\Lambda^*$ \cite{kstar200,baryon}
resonances. Now, with the addition of the $d$+Au measurement we
can gain insight into the re-scattering processes in $p+p$ and
Au+Au collisions.

In this paper, we present the first measurements of
$\rho$(770)$^0$, $K^*$(892), $\Delta$(1232)$^{++}$,
$\Sigma$(1385), and $\Lambda$(1520) in $d$+Au collisions at
$\sqrt{s_{_{NN}}}$ = 200 GeV  at RHIC reconstructed via their
hadronic decay channels using the STAR detector. The
$\rho(770)^0$, $K^*(892)$, $\Delta(1232)^{++}$, and $\Sigma(1385)$
masses are presented as a function of $p_T$ in $d$+Au collisions
and the $\rho^0$ and $\Delta^{++}$ masses are compared to the
measurements in $p+p$ collisions. The $p_T$ spectra of these
resonances are presented for different centralities in $d$+Au
collisions. The $\langle p_T \rangle$ of resonances measured in
minimum bias collisions are compared to the $\langle p_T \rangle$
of $\pi$, $K$, and $\overline{p}$. The $\rho^0/\pi^-$, $K^*/K^-$,
$\Delta^{++}/p$, $\Sigma(1385)/\Lambda$, and
$\Lambda(1520)/\Lambda$ ratios in $d$+Au and minimum bias $p+p$
collisions are compared. The nuclear modification factors $R_{dAu}$
of these resonances are compared to the $R_{dAu}$ of charged
hadrons and the Cronin (initial state multiple scattering)
enhancement \cite{cronin} is discussed.

\section{Experiment}

We present measurements of resonances via their hadronic decay
channels (see Table \ref{tab:BR}) in $d$+Au collisions at
$\sqrt{s_{_{NN}}}$ = 200 GeV using the Time Projection Chamber
(TPC) \cite{tpc}, which is the primary tracking device of the STAR
experiment.

A minimum bias trigger was defined by requiring that at least one
beam-rapidity neutron impinges on the Zero Degree Calorimeter
(ZDC) \cite{zdc} in the Au beam direction. The measured minimum
bias cross-section amounts to $95 \pm 3\%$ of the total $d$+Au
geometric cross-section. Charged particle multiplicity within the
pseudorapidity interval $-3.8 < \eta < -2.8$ was measured in a
forward TPC (FTPC) \cite{ftpc} in the Au beam direction and served
as the basis of our $d$+Au centrality tagging scheme, as described
elsewhere \cite{cent}. The $d$+Au centrality definition consists
of three event centrality classes; 0-20$\%$, 20-40$\%$, and
40-100$\%$ of the total $d$+Au cross-section \cite{bin}. The
analysis of different centralities was restricted to events with a
primary vertex within 50 cm of the center of the TPC along the
beam direction to ensure uniform acceptance in the $\eta$ range
studied. For the minimum bias events, the FTPC was not used, so
events with a primary vertex within 100 cm were accepted, still
maintaining a uniform acceptance in the $\eta$ range studied. In
order to improve the statistics in the case of the $K^*$(892),
events with a primary vertex within 75 cm were accepted for the
centrality studies. The difference between using 50 and 75 cm as
the primary vertex cut was taken into account in the systematic
errors. The same primary vertex cut was used for the minimum bias
events. In order to improve statistics in the case of the
$\Lambda(1520)$, events with a primary vertex within 70 cm were
accepted for centrality selected minimum bias events. A summary of
the relevant data-sets is given in Table \ref{tab:dataset}.

\begin{table}
\caption{\label{tab:BR}The resonances measured in $d$+Au
collisions, their corresponding hadronic decay channels, branching
ratios, and lifetimes \cite{pdg}.}
\begin{ruledtabular}
\begin{tabular}{cccc}
Resonance & Decay Channel & B.R. & $c\tau$ \\
\hline
    $\rho^0(770)$ & $\pi^+ \pi^-$ & $\sim 100\%$ & 1.3 fm \\
    $K^*(892)^0$ & $K^+\pi^-$ & $\sim 66.7\%$ & 3.9 fm \\
    $\overline{K}^*(892)^0$ & $K^-\pi^+$ & $\sim 66.7\%$ & 4 fm \\
    $K^*(892)^{\pm}$ & $K_{S}^{0}\pi^{\pm}$ & $\sim 66.7\%$ & 4 fm \\
    $\Delta(1232)^{++}$ & $p\pi^+$ & $\sim 100\%$ & 1.6 fm \\
    $\overline{\Delta}(1232)^{--}$ & $\overline{p}\pi^-$ & $\sim 100\%$ & 1.6 fm \\
    $\Sigma(1385)^{+}$ & $\Lambda\pi^{+}$ & $\sim 87\%$ & 5.5 fm \\
    $\overline{\Sigma}(1385)^{-}$ & $\overline{\Lambda}\pi^{-}$ & $\sim 87\%$ & 5.5 fm \\
    $\Sigma(1385)^{-}$ & $\Lambda\pi^{-}$ & $\sim 87\%$ & 5.0 fm \\
    $\overline{\Sigma}(1385)^{+}$ & $\overline{\Lambda}\pi^{+}$ & $\sim 87\%$ & 5.0 fm \\
    $\Lambda(1520)$ & $pK^-$ & $\sim 22.5\%$ & 12.6 fm\\
    $\overline{\Lambda}(1520)$ & $\overline{p}K^+$ & $\sim 22.5\%$ & 12.6 fm\\
\end{tabular}
\end{ruledtabular}
\end{table}

\begin{table}
\caption{\label{tab:dataset}The data-set for each centrality used
in the analysis of resonances in $d$+Au collisions.}
\begin{ruledtabular}
\begin{tabular}{cccc}
Centrality & Number of & Primary & Resonance\\
& Events & Vertex (cm) &\\
 \hline
    Minimum Bias & $\sim 16 \times 10^6$ & $\pm$ 100 & $\rho^0$, $\Sigma^*$\\
    Minimum Bias & $\sim 15 \times 10^6$ & $\pm$ 75 &  $K^*$\\
    Minimum Bias & $\sim 14 \times 10^6$ & $\pm$ 70 & $\Lambda^*$\\
    Minimum Bias & $\sim 11.6 \times 10^6$ & $\pm$ 50 & $\Delta^{++}$\\
\end{tabular}
\end{ruledtabular}
\end{table}

We also present measurements in $p+p$ collisions of the
$\Delta^{++}$ where a minimum bias trigger was defined using
coincidences between two beam-beam counters that measure the
charged particle multiplicity at forward pseudorapidities ($3.3 <
|\eta| < 5.0$). In this case, $\sim 6 \times 10^6$ events were
used, where only events with a primary vertex within $\pm$50 cm
were accepted.

\begin{figure}[tbp]
\includegraphics[width=0.47\textwidth]{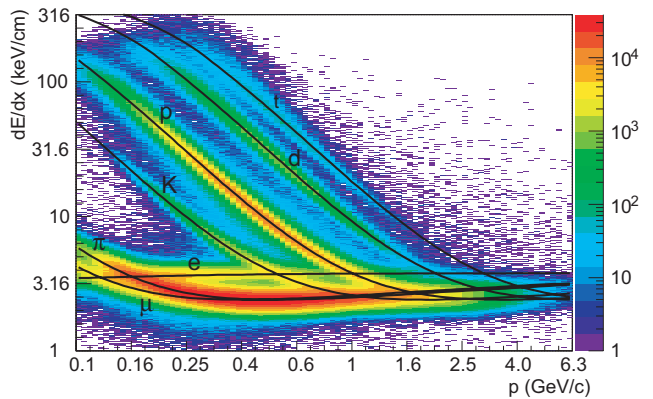}
\caption{\label{fig:dEdx}(Color online) $dE/dx$  for charged
particles vs. momentum measured by the TPC in $d$+Au collisions.
The curves are the Bichsel function \cite{bic} for different
particle species.}
\end{figure}

In addition to momentum information, the TPC provides particle
identification for charged particles by measuring their ionization
energy loss ($dE/dx$). Fig. \ref{fig:dEdx} shows $dE/dx$ as a
function of momentum ($p$) measured in the TPC. The different
bands presented in Fig. \ref{fig:dEdx} represent Bichsel
distributions folded with the experimental resolutions and
correspond to different particle species \cite{bic}. Charged pions
and kaons can be separated in momenta up to about 0.75 GeV/$c$,
while (anti-)protons can be identified for momenta of up to about 1.1
GeV/$c$. In Fig. \ref{fig:dEdx}, the Bichsel function \cite{bic}
is used instead of the traditional Bethe-Bloch parametrization
\cite{pdg} in order to improve particle identification. To
quantitatively describe the particle identification, the variable
$N_{\sigma\pi}$, which expresses energy loss in the units of the
standard deviation of a Gaussian formed by the logarithm of
truncated energy loss, is defined (in this case for pions) as:
\begin{equation}
N_{\sigma\pi}=\frac{1}{\sigma_{dE/dx}(L_{\text{TPC}})} \log
\frac{dE/dx_{\text{measured}}}{\langle dE/dx\rangle_\pi},
\label{eq:NSigma}
\end{equation}
where $dE/dx_{\text{measured}}$ is the measured energy loss for a
track, $\langle dE/dx\rangle_\pi$ is the expected mean energy loss
for charged pions with a given momentum, and
$\sigma_{dE/dx}(L_{\text{TPC}})$ is the $dE/dx$ resolution that
depends on the track length in the TPC that is used in the $dE/dx$
measurement. For $L_{\text{TPC}}$ equal to 72 cm, corresponding to
a 90$^0$ angle with the beam axis, the resolution is 8.1$\%$. In
the case of charged kaon and charged proton identification,
similar definitions of $N_{\sigma K}$ and $N_{\sigma p}$ can be
obtained. In order to quantitatively select on charged pions,
kaons, and protons, specific analysis cuts, described later, are
then applied to the variables $N_{\sigma\pi}$, $N_{\sigma K}$, and
$N_{\sigma p}$.

\section{Particle Selection}

In all cases, particles and anti-particles are combined in order
to improve statistics. In the following, the term $K^{*0}$ stands
for $K^{*0}$ or $\overline{K}^{*0}$, and the term $K^*$ stands for
$K^{*0}$, $\overline{K}^{*0}$ or $K^{*\pm}$, unless otherwise
specified. The term $\Delta^{++}$ stands for $\Delta^{++}$ or
$\overline{\Delta}^{--}$, the term $\Sigma^*$ stands for
$\Sigma(1385)^{+}$, $\Sigma(1385)^{-}$,
$\overline{\Sigma}(1385)^{-}$ or $\overline{\Sigma}(1385)^{+}$,
and the term $\Lambda^*$ stands for $\Lambda(1520)$ or
$\overline{\Lambda}(1520)$, unless otherwise specified.

As these studied resonances decay in such short times that the
daughters seem to originate from the interaction point, only
charged pion, kaon, and proton candidates whose distance of
closest approach to the primary interaction vertex was less than 3
cm were selected. Such candidate tracks are referred to as primary
tracks. In order to avoid the acceptance drop in the high $\eta$
range, all track candidates were required to have $|\eta| < 0.8$.
For all candidates, in order to avoid selecting split tracks, a
cut on the ratio of the number of TPC track fit points and the
maximum possible points was required. In addition, a minimum $p_T$
cut was applied to maintain reasonable momentum resolution.

\begin{figure}[tbp]
\includegraphics[width=0.45\textwidth]{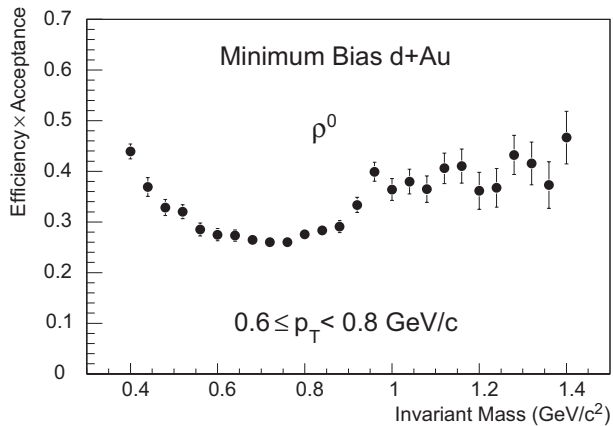}
\caption{\label{fig:accrho} The $\rho^{0}$ reconstruction
efficiency$\times$detector acceptance as a function of the
invariant mass for minimum bias $d$+Au. The error shown is due to
the available statistics in the simulation.}
\end{figure}

In the case of the $\rho^0$, a series of cuts was applied to the
charged pion candidates in order to ensure track fit quality and
good $dE/dx$ resolution. A compilation of the cuts used in the
$\rho^0$ analysis is given in Table \ref{table:cuts} and the
$\rho^0$ correction factor (reconstruction efficiency multiplied
by the detector acceptance) as a function of invariant mass for a
particular $p_T$ bin is depicted in Fig. \ref{fig:accrho}. In
general, the correction factor increases as a function of
transverse momentum. The fact that the correction factor is larger
at low values of $M_{\pi\pi}$ and larger values of $p_T$ is simply
due to kinematics. In the case of wide resonances, such as the
$\rho^0$ and the $\Delta^{++}$, the correction factor depends on
the invariant mass for each $p_T$ interval that is being analyzed.
In this case, the correction is applied as a function of the
invariant mass for each $p_T$ bin. In the case of narrow
resonances, such as the $K^*$, $\Sigma^*$, and $\Lambda^*$, the
correction factor is dependent only on the $p_T$ bin being
analyzed. Therefore, the correction is performed as a function of
$p_T$ only.

\begin{table*}
\caption{List of track cuts for charged kaons and charged pions
  and
  topological cuts for neutral kaons used in the $\rho^0$ and $K^{*}$ analyses in $d$+Au
  collisions. $decayLength$ is the decay length, $dcaDaughters$
  is the distance of closest approach between the daughters, $dcaV0PrmVx$ is
  the distance of closest approach between the reconstructed $K_{S}^{0}$ momentum
  vector and the primary interaction vertex, $dcaPosPrmVx$ is the distance of
  closest approach between the positively charged granddaughter and the primary vertex,
  $dcaNegPrmVx$ is the distance of closest approach between the negatively charged
  granddaughter and the primary vertex, $M_{K_S^0}$ is the $K_S^0$ invariant mass in
  GeV/$c^2$, $NFitPnts$ is the number of fit points of a track in the TPC, $NTpcHits$ is the
  number of hits of a track in the TPC, $MaxPnts$ is the number of maximum possible
  points of a track in the TPC, and $DCA$ is the distance of closest approach to the
  primary interaction point. The Normalization Region corresponds to the interval in which
  the invariant mass and the background reference distributions are normalized.}\label{table:cuts}
\centering
\begin{ruledtabular}
\begin{tabular}{c c c c c}
    & $\rho^0$ & $K^{*0}$ & \multicolumn{2}{c}{$K^{*\pm}$}\\
    Cuts\\
    & & & Daughter $\pi^{\pm}$ & $K_{S}^{0}$ \\ \hline
    $N_{\sigma K}$& & (-2.0, 2.0) &  & $decayLength > $ 2.0 cm \\
    $N_{\sigma\pi}$ & (-3.0, 3.0)& (-3.0, 3.0) & (-2.0, 2.0) & $dcaDaughters < $ 1.0 cm \\
    Kaon $p$ (GeV/$c$) & $>$ 0.2& (0.2, 0.7) & & $dcaV0PrmVx <$ 1.0 cm \\
    Kaon $p_{T}$(GeV/$c$) & $>$ 0.2 &(0.2, 0.7) & & $dcaPosPrmVx >$ 0.5 cm\\
    Pion $p$(GeV/$c$) & $>$ 0.2 & (0.2, 10.0) &  (0.2, 10.0) & $dcaNegPrmVx >$ 0.5 cm\\
    Pion $p_{T}$ (GeV/$c$) & $>$ 0.2  & (0.2, 10.0) & (0.2, 10.0) & $M_{K_S^0}$ (GeV/$c^2$): (0.48, 0.51)\\
    $NFitPnts$ & $>$ 15 & $>$ 15 & $>$ 15 & $\pi^{+}$: $NTpcHits >$ 15\\
    $NFitPnts/MaxPnts$ & $>$ 0.55  & $>$ 0.55  & $>$ 0.55 &$\pi^{-}$: $NTpcHits >$ 15 \\
    Kaon and pion $\eta$ & $|\eta| <$ 0.8  & $|\eta| <$ 0.8  & $|\eta| <$ 0.8 & $\pi^{+}$: $p>$ 0.2 GeV/c\\
    $DCA$ (cm) & $<$ 3.0 & $<$ 3.0 & $<$ 3.0 & $\pi^{-}$: $p >$ 0.2 GeV/$c$ \\
    Mass Normalization Region (GeV/$c^2$)& (1.5,2.5) & & \\ \hline
    Pair $y$ & $|y| <$ 0.5 & $|y| <$ 0.5 & $|y| <$ 1.0 & \\
\end{tabular}
\end{ruledtabular}
\end{table*}

\begin{figure}[tbp]
\includegraphics[width=0.45\textwidth]{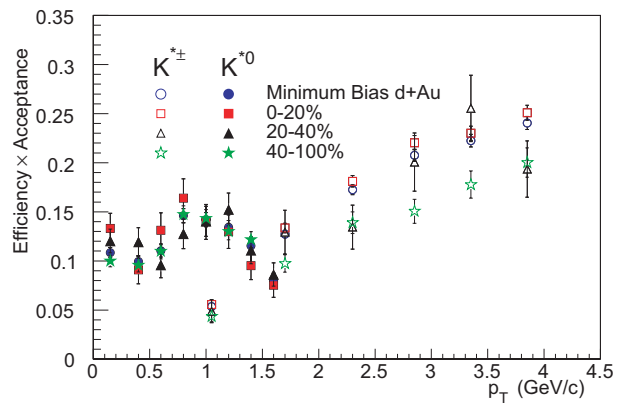}
\caption{\label{fig:acckstar} (Color online) The $K^{*0}$ and
$K^{\pm}$ reconstruction efficiency$\times$detector acceptance as
a function of $p_T$ for minimum bias $d$+Au and three different
centralities.}
\end{figure}

For the $K^*$ analysis, charged kaon candidates were selected by
requiring $|N_{\sigma K}| < 2$ while a looser cut $|N_{\sigma\pi}|
< 3$ was applied to select the charged pion candidates to maximize
statistics for the $K^{*}$ analysis. Such $N_{\sigma}$ cuts do not
unambiguously select kaons and pions, but do help to reduce the
background significantly. The background was reduced further by
selecting only kaon candidates with $p < 0.7$ GeV/$c$. This
momentum cut ensures clearer identification by minimizing
contamination from misidentified correlated pairs and thus the
systematic uncertainties \cite{kstar200}. The $K^{*\pm}$ first
undergoes a strong decay to produce a $K_S^0$ and a charged pion
hereafter labelled as the $K^{*\pm}$ daughter pion. Then, the
produced $K_S^0$ decays weakly into a $\pi^+\pi^-$ pair with
$c\tau = $ 2.67 cm. The oppositely charged pions from the $K_S^0$
decay are called the $K^{*\pm}$ granddaughter pions. The charged
daughter pion candidates were selected from primary track samples
and the $K_S^0$ candidates were selected through their decay
topology \cite{top1,top2}. The procedure is briefly outlined
below. The granddaughter charged pion candidates were selected
from tracks that do not originate from the primary collision
vertex. Oppositely charged candidates were then paired to form
neutral decay vertices. When the $K^0_S$ candidate was paired with
the daughter pion to reconstruct the charged $K^*$, tracks were
checked to avoid double counting among the three tracks used. Cuts
were applied to the daughter and granddaughter candidates to
ensure track fit quality and good $dE/dx$ resolution and to reduce
the combinatorial background in the $K^0_S$ invariant mass
distribution. All the cuts used in this $K^*$ analysis are
summarized in Table \ref{table:cuts} and the $K^*$ reconstruction
correction factors are shown in Fig. \ref{fig:acckstar}.

\begin{figure}[tbp]
\includegraphics[width=0.45\textwidth]{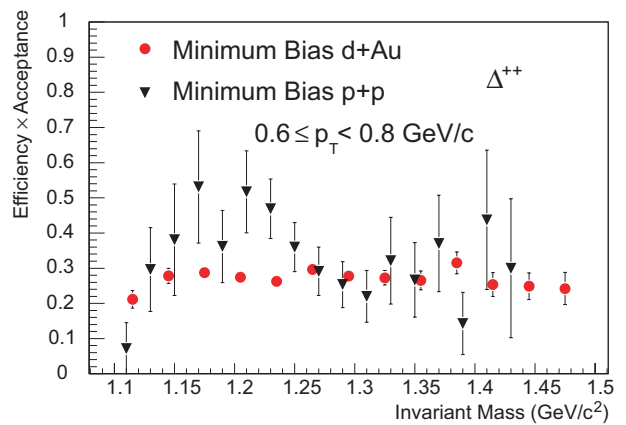}
\caption{\label{fig:accdelta} (Color online) The $\Delta^{++}$
reconstruction efficiency$\times$detector acceptance as a function
of the invariant mass for minimum bias $d$+Au and $p+p$. The error
shown is due to the available statistics in the simulation.}
\end{figure}

\begin{figure}[tbp]
\includegraphics[width=0.45\textwidth]{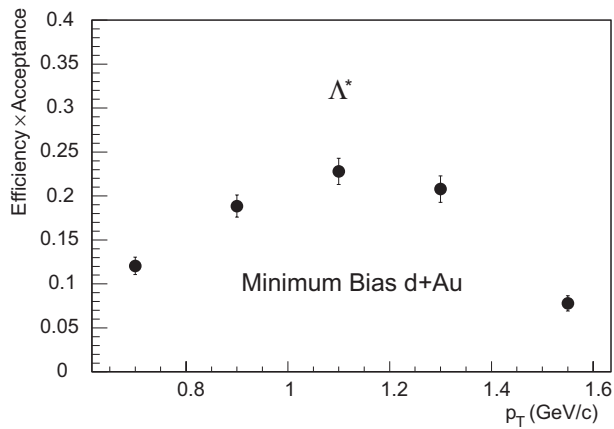}
\caption{\label{fig:acclambda} The $\Lambda^*$ reconstruction
efficiency$\times$detector acceptance as a function of $p_T$ for
minimum bias $d$+Au. The error shown is due to the available
statistics in the simulation.}
\end{figure}

\begin{figure}[tbp]
\includegraphics[width=0.45\textwidth]{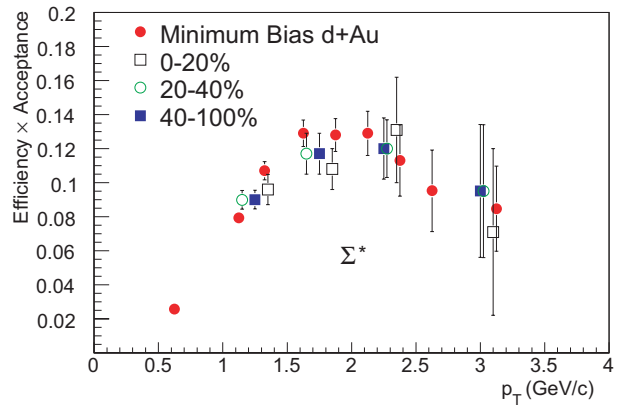}
\caption{\label{fig:accsigma} The $\Sigma^{*\pm}$ reconstruction
efficiency$\times$detector acceptance as a function of $p_T$ for
minimum bias $d$+Au and three different centralities. The error
shown is due to the available statistics in the simulation.}
\end{figure}

The cuts applied to the $\Delta^{++}$ and $\Lambda^*$ decay
daughters were the same as described above for the $\rho^0$ and
$K^{*0}$, and their respective values are given in Table
\ref{table:cuts1}. We also present the $\Delta^{++}$ measured in
$p+p$ collisions and the cuts and their respective values applied
to the decay daughters are the same as the ones used in the $d$+Au
analysis and shown in Table \ref{table:cuts1}. Similarly to the
$K^{*\pm}$, the $\Sigma^{*\pm}$ first undergoes a strong decay to
produce a $\Lambda$, which subsequently decays weakly into $\pi^-
p$ with a $c\tau = $ 7.89 cm. The cuts applied to the
$\Sigma^{*\pm}$ decay daughters and granddaughters are the same as
mentioned for the $K^{*\pm}$ and the values are shown in Table
\ref{table:cuts1}. The $\Delta^{++}$, $\Lambda^*$, $\Sigma^{*\pm}$
reconstruction correction factors are shown in Figs.
\ref{fig:accdelta}, \ref{fig:acclambda}, and \ref{fig:accsigma},
respectively.

\begin{table*}
\caption{List of track cuts for charged kaons, charged pions,
  charged protons and
  topological cuts for lambdas used in the $\Delta^{++}, \Lambda^*$, and $\Sigma^{*\pm}$
  analyses in $d$+Au
  collisions. $decayLength$ is the decay length, $dcaDaughters$
  is the distance of closest approach between the daughters, $dcaV0PrmVx$ is
  the distance of closest approach between the reconstructed $\Lambda$ momentum
  vector and the primary interaction vertex, $dcaPosPrmVx$ is the distance of
  closest approach between the positively charged granddaughter and the primary vertex,
  $dcaNegPrmVx$ is the distance of closest approach between the negatively charged
  granddaughter and the primary vertex, $M_{\Lambda}$ is the $\Lambda$ invariant mass in
  GeV/$c^2$, $NFitPnts$ is the number of fit points of a track in the TPC, $NTpcHits$ is the
  number of hits of a track in the TPC, $MaxPnts$ is the number of maximum possible
  points of a track in the TPC, $\theta^*$ is the angle in the center-of-mass of one decay particle
  with respect to the mother particle, and $DCA$ is the distance of closest approach to the
  primary interaction point. The Normalization Region corresponds to the interval in which
  the invariant mass and the background reference distributions are normalized.}\label{table:cuts1}
\centering
\begin{ruledtabular}
\begin{tabular}{c c c c c}
    & $\Delta^{++}$ & $\Lambda^*$ & \multicolumn{2}{c}{$\Sigma^{*\pm}$}\\
    Cuts\\

    & & & Daughter $\pi^{\pm}$ & $\Lambda$ \\ \hline

    $N_{\sigma K}$& & (-2.0, 2.5) &  &  \\

    $N_{\sigma p}$& (-2.0, 2.0) & (-2.0, 2.5) &  & $decayLength$ (cm): (5.0,30.0) \\

    $N_{\sigma\pi}$ & (-2.0, 2.0) &  & (-3.0, 3.0) & $dcaDaughters < $ 1.0 cm \\

    Kaon $p$ (GeV/$c$) & & (0.2,0.8) & & \\

    Kaon $p_{T}$(GeV/$c$) &  & & & \\

    Proton $p$ (GeV/$c$) & (0.3,1.1) & (0.2,1.0) & & $dcaV0PrmVx <$ 1.1 cm \\

    Proton $p_{T}$(GeV/$c$) & (0.3,1.1) & & & $dcaPosPrmVx >$ 0.9 cm\\

    Pion $p$(GeV/$c$) & (0.1,0.6) & &  (0.15, 1.5) & $dcaNegPrmVx >$ 2.5 cm\\

    Pion $p_{T}$ (GeV/$c$) & (0.1,0.6) &  &  & $M_{\Lambda}$ (GeV/$c^2$): (1.11, 1.12)\\

    & Proton $p$ $>$ Pion $p$ & & &\\

    $NFitPnts$ & $>$ 15 & $>$ 20 & $>$ 15 & $p$: $NTpcHits >$ 15\\

    $NFitPnts/MaxPnts$ & $>$ 0.55  & $>$ 0.51  & $>$ 0.55 &$\pi^{-}$: $NTpcHits >$ 15 \\

    Proton and pion $\eta$ & $|\eta| <$ 0.8  &  & $|\eta| <$ 1.5 & $p$: $p>$ 0.1 GeV/c\\


    $DCA$ (cm) & $<$ 3.0 & $<$ 1.5 & $<$ 1.5 & $\pi^{-}$: $p >$ 0.1 GeV/$c$ \\

    $\cos \theta^*$ &  & (-0.8,0.8) & & \\

    Mass Normalization  Region (GeV/$c^2$) & See text & (1.55-1.8) & (1.45-2.0) &\\\hline

    Pair $y$ & $|y| <$ 0.5 & $|y| <$ 0.5 & $|y| <$ 0.75 & \\
\end{tabular}
\end{ruledtabular}
\end{table*}

\section{Analysis and Results}

The $\rho^0$ measurement was performed by calculating the
invariant mass for each $\pi^+\pi^-$ pair in an event which passed
the cuts. The resulting invariant mass distribution was then
compared to a reference distribution calculated from the geometric
mean of the invariant mass distributions obtained from uncorrelated
like-sign pion pairs from the same events \cite{kstar200}. The
$\pi^+\pi^-$ invariant mass distribution ($M_{\pi\pi}$) and the
like-sign reference distribution were normalized to each other
between 1.5 $ \leq M_{\pi\pi} \leq$ 2.5 GeV/$c^2$.

The $K^{*}$, $\Delta^{++}$, $\Sigma^*$, and $\Lambda^*$
measurements were performed using the mixed-event technique
\cite{kstar200}, where the reference background distribution is
built with uncorrelated unlike-sign kaons and pions, protons and
pions, lambdas and pions and protons and kaons from different
events, respectively. The background is normalized over a wide
kinematic range (see Tables \ref{table:cuts} and
\ref{table:cuts1}) and then subtracted from the corresponding
invariant mass distribution.

\subsection{Masses and Widths}

The mass and width of resonances have been of
great interest because of their possible modification in the medium produced
in heavy-ion collisions \cite{rapp3}. It is interesting to study how the resonance
masses and widths behave in $d$+Au collisions.

\begin{figure}[tbp]
\includegraphics[width=0.45\textwidth]{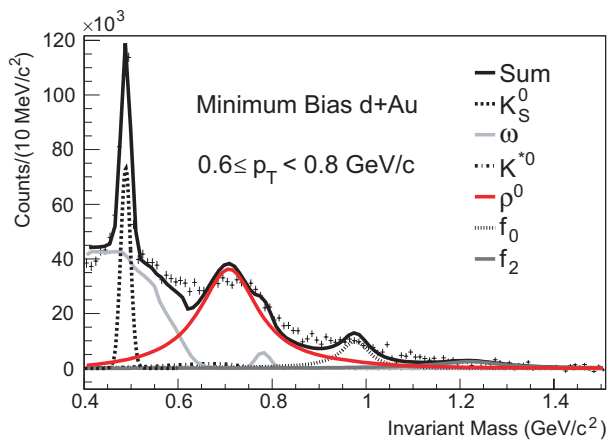}
\caption{\label{fig:cocktail}(Color online) The raw $\pi^+\pi^-$
invariant mass distribution  at midrapidity after subtraction of
the like-sign reference distribution for minimum bias $d$+Au
collisions with the hadronic cocktail fit. }
\end{figure}

The corresponding $\pi^+\pi^-$ raw invariant mass distribution
after the like-sign background subtraction for minimum bias $d$+Au
collisions at midrapidity ($|y| < 0.5$) for a particular $p_T$ bin
is shown in Fig. \ref{fig:cocktail}. The solid black line in
Fig.~\ref{fig:cocktail} is the sum of all the well defined
contributions to the $M_{\pi\pi}$ distribution (hadronic cocktail) \cite{rho}.
The $K_S^0$ was fit with a Gaussian function. The $\omega$ and
$K^{\ast}(892)^{0}$ shapes were obtained from the HIJING event
generator \cite{wan}, with the kaon being misidentified as a pion
in the case of the $K^{\ast 0}$. The $\rho^0(770)$, the $f_0(980)$
and the $f_2(1270)$ were fit to a BW$\times$PS function where BW
is the relativistic Breit-Wigner function \cite{rhop}
\begin{equation}
\mathrm{BW} = \frac{M_{\pi\pi}M_0\Gamma}{[(M_0^2 \!-\!
M_{\pi\pi}^2)^2 \!+\! M_0^2\Gamma^2]}
\label{bwstar}
\end{equation}
and PS is the Boltzmann factor \cite{rapp4,shuryak,pbm,kolb}
\begin{equation}
\mathrm{PS} = \frac{M_{\pi\pi}}{\sqrt{M_{\pi\pi}^2 + p_T^2}}
\times \exp(-\frac{\sqrt{M_{\pi\pi}^2 + p_T^2}}{T})
\end{equation}
to account for phase space. Here, $T$ is the temperature parameter
at which the resonance is emitted \cite{shuryak} and
\begin{equation}
\Gamma = \Gamma_0 \times \frac{M_0}{M_{\pi\pi}} \times
[\frac{(M_{\pi\pi}^2 - 4m_\pi^2)}{(M_0^2 - 4m_\pi^2)}]^{(2\ell +
1)/2}
\end{equation}
is the width \cite{rhop}, which changes as a function of momentum
due to reconstruction effects. Here, $M_0$ and $\ell$ are the
resonance mass and spin, respectively. The masses of $K_S^0$,
$\rho^0$, $f_0$, and $f_2$ were free parameters in the fit, and
the widths of $f_0$ and $f_2$ were fixed according to \cite{pdg}.
The uncorrected yields of $K_S^0$, $\rho^0$, $\omega$, $f_0$, and
$f_2$ were free parameters in the fit while the $K^{\ast 0}$
fraction was fixed according to the $K^{\ast}(892)^{0}
\!\rightarrow\! \pi K$ measurement, where the amount of
contamination was determined using a detailed simulation of the
TPC response using GEANT \cite{detectorres}. The $\rho^0$,
$\omega$, $K^{\ast 0}$, $f_0$, and $f_2$ distributions were
corrected for the detector acceptance and efficiency determined
from simulation. The signal to background ratio before subtraction
is about 1/100.

The cocktail fit does not reproduce the $\pi^+\pi^-$ raw invariant
mass distribution at $\sim$600 and $\sim$850 MeV/$c^2$,
respectively. This is understood to be due to other contributions
to the hadronic cocktail aside from what was described above, e.g.
the $f_0(600)$ that is not very well established \cite{pdg}.
The $\omega$ yield in the hadronic cocktail fits may account for
some of these contributions and may cause the apparent decrease in
the $\rho^0/\omega$ ratio between minimum bias $p+p$ and
peripheral Au+Au interactions.

\begin{figure}[tbp]
\includegraphics[width=0.45\textwidth]{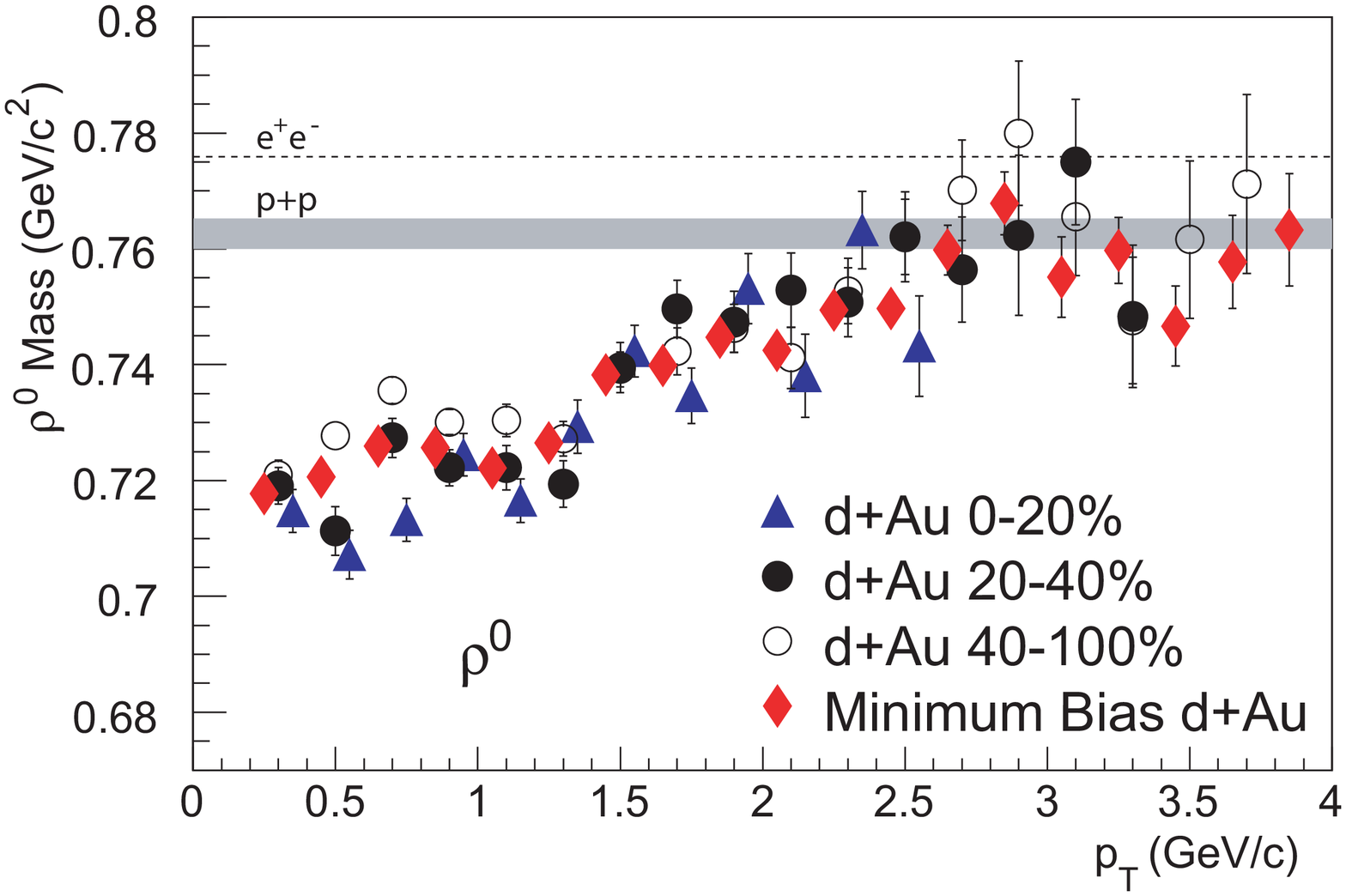}
\includegraphics[width=0.45\textwidth]{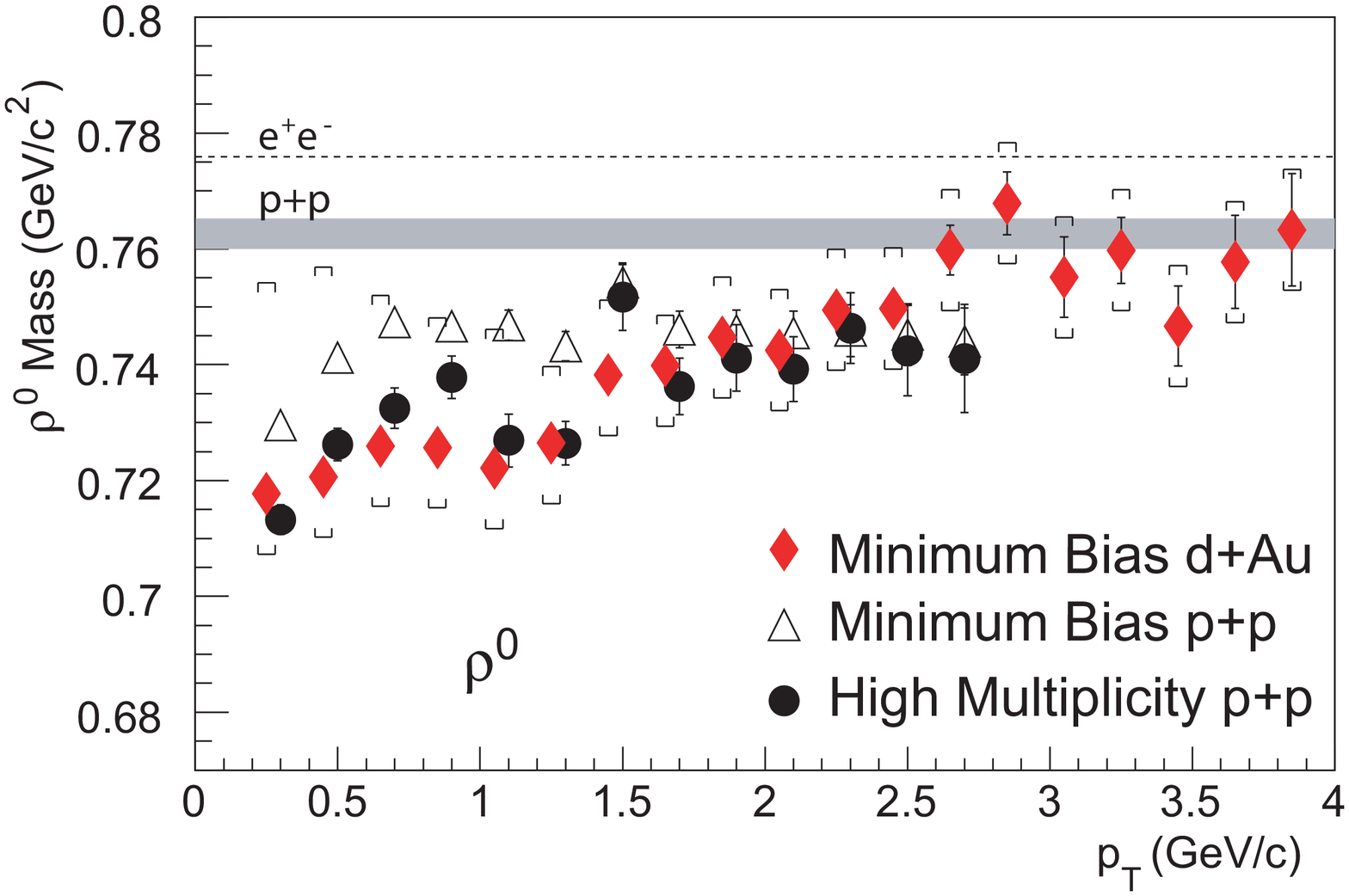}
\caption{(Color online) The $\rho^0$ mass as a function of $p_T$
at $|y| < 0.5$ for minimum bias $d$+Au interactions and 0-20$\%$,
20-40$\%$, and 40-100$\%$ of the total $d$+Au cross-section (upper
panel). The errors shown are statistical only. The comparison of
the $\rho^0$ mass as a function of $p_T$ at $|y| < 0.5$ measured
in minimum bias $d$+Au, $p+p$, and high multiplicity $p+p$
\cite{rho} interactions (lower panel). The brackets indicate the
systematic uncertainty and it is shown only for the minimum bias
$d$+Au measurement for clarity. The diamonds have been shifted to
lower values on the abscissa by 100 MeV/$c$ in $p_T$ for clarity.
}\label{fig:massrho}
\end{figure}

The $\rho^0$ mass is shown as a function of $p_T$ in Fig.
\ref{fig:massrho} for minimum bias $d$+Au interactions and
0-20$\%$, 20-40$\%$, and 40-100$\%$ of the total $d$+Au
cross-section. A mass shift of $\sim$50 MeV/$c^2$ is observed at
low $p_T$. The $\rho^0$ width was fixed at $\Gamma_0$ = 160
MeV$/c^2$, consistent with folding the $\rho^0$ natural width
(150.9 $\pm$ 2.0 MeV$/c^2$ \cite{pdg}) with the intrinsic
resolution of the detector ($\delta p_T/p_T =
0.005(1+p_T)$) \cite{detectorres}. The temperature parameter used
in the PS factor was $T$ = 160 MeV, which was also used in the
$p+p$ analysis \cite{rho}. In Fig. \ref{fig:massrho}, only the
systematic uncertainty for the minimum bias $d$+Au measurement is
depicted for clarity. However, the systematic uncertainty for the
other $d$+Au centrality measurements are similar, if not less than
the systematic uncertainty for the minimum bias $d$+Au
measurement.

The $\rho^0$ mass at $|y| < $ 0.5 for minimum bias and three
different centralities in $d$+Au collisions at $\sqrt{s_{NN}}$ =
200 GeV increases as a function of $p_T$ and is systematically
lower than the value reported by NA27 at CERN-LEBC-EHS
\cite{rhopp}. This experiment measured the $\rho^0$ in minimum
bias $p+p$ collisions at $\sqrt{s} = $ 27.5 GeV for $x_F > 0$,
where $x_F$ is the ratio between the longitudinal momentum and the
maximum momentum of the meson. In Fig. \ref{fig:massrho}, the
shaded areas indicate the $\rho^0$ mass measured in $p+p$
collisions (762.6 $\pm$ 2.6 MeV/$c^2$) by NA27 \cite{rhopp} and
the dashed lines represent the average of the $\rho^0$ mass
measured in $e^+e^-$ (775.6 $\pm$ 0.5 MeV/$c^2$) \cite{pdg}. The
$\rho^0$ mass measured in 0-20$\%$ of the total $d$+Au
cross-section is slightly lower than the mass measured in the most
peripheral centrality class. The masses measured in minimum bias
$d$+Au, $p+p$ \cite{rho}, and high multiplicity $p+p$ \cite{rho}
interactions are compared in Fig. \ref{fig:massrho}. The
comparison shows that the $\rho^0$ mass measured in minimum bias
$d$+Au and high multiplicity $p+p$ interactions are comparable. A
mass shift of $\sim$70 MeV/$c^2$ was also measured in Au+Au
collisions \cite{rho}. Dynamical interactions with the surrounding
matter, interference between various $\pi^+\pi^-$ scattering
channels, phase space distortions due to the re-scattering of
pions forming $\rho^0$, and Bose-Einstein correlations between
$\rho^0$ decay daughters and pions in the surrounding matter were
previously given as the possible explanations \cite{rho}. It has
been proposed \cite{fac} that the mass shift observed in $p+p$
collisions is due to $\pi\pi$ re-scattering, which requires no
medium. Since one also does not expect a medium to be formed in
$d$+Au collisions, if dynamical interactions are also the
explanation for the mass shift, then the re-scattering of the
$\rho^0$ with the surrounding particles must exist. We also
observe that the $\rho^0$ mass is not modified at high $p_T$.

\begin{figure}[tbp]
\begin{center}
\resizebox{0.48\textwidth}{!}{%
  \includegraphics{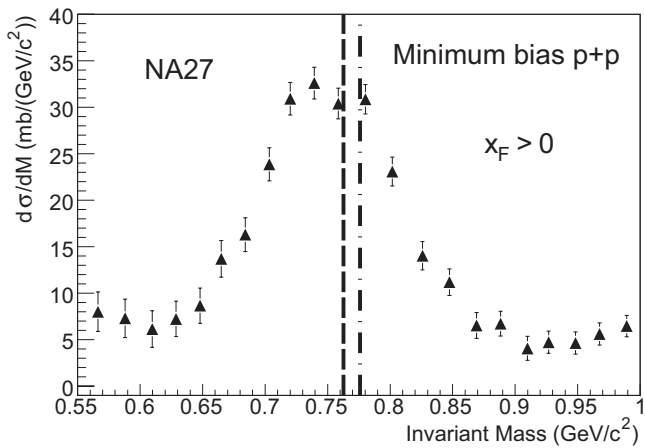}
  }
\caption{The invariant $\pi^+\pi^-$ mass distribution after
background subtraction for minimum bias $p+p$ collisions measured
by NA27. For details see elsewhere \cite{rhopp}. The vertical
dash-dotted line represent the average of the $\rho^0$ mass 775.8 $\pm$
0.5 MeV/$c^2$ measured in $e^+e^-$ \cite{pdg}. The vertical
dashed line is the $\rho^0$ mass 762.6 $\pm$ 2.6 MeV/$c^2$
reported by NA27 \cite{rhopp}. }\label{fig:massNA27}
\end{center}
\end{figure}

NA27 measured the $\rho^0$ in minimum bias $p+p$ at $\sqrt{s} = $
27.5 GeV for $x_F > 0$ and reported a mass of 762.6 $\pm$ 2.6
MeV/$c^2$ \cite{rhopp}. The invariant $\pi^+\pi^-$ mass
distribution after subtraction of the mixed-event reference
distribution is shown in Fig. \ref{fig:massNA27}. The vertical
dash-dotted line represents the average of the $\rho^0$ mass 775.8
$\pm$ 0.5 MeV/$c^2$ measured in $e^+e^-$ collisions \cite{pdg}.
The vertical dashed line, which accounts for the phase space, is
the $\rho^0$ mass reported by NA27 (762.6 $\pm$ 2.6 MeV/$c^2$)
\cite{rhopp}. As shown in Fig. \ref{fig:massNA27}, the position of
the $\rho^0$ peak is shifted by $\sim$ 30 MeV/$c^2$ compared to
the $\rho^0$ mass in the vacuum 775.8 $\pm$ 0.5 MeV/$c^2$
\cite{pdg}.

NA27 obtained the $\rho^0$ mass by fitting the invariant
$\pi^+\pi^-$ mass distribution by the (BW$\times$PS + BG)
function, where in this analysis, the phase space function used is
the same as the combinatorial background (BG). NA27 reported a
mass of 762.6 $\pm$ 2.6 MeV/$c^2$, which is $\sim$ 10 MeV/$c^2$
lower than the $\rho^0$ mass in the vacuum. Ideally, the PS factor
should have accounted for the shift on the $\rho^0$ peak, and the
mass obtained from the fit should have agreed with the $\rho^0$
mass in the vacuum. However, just like in the STAR measurement,
this was not the case, since the phase space did not account for
the mass shift on the position of the $\rho^0$ peak.

At the CERN-LEP accelerator, OPAL, ALEPH and DELPHI measured the
$\rho^0$ in inclusive $e^+e^-$ reactions at $\sqrt{s} = $ 90 GeV
\cite{act,laf,ack,bus}. OPAL reported a shift in the position of
the $\rho^0$ peak by $\sim$ 70 MeV/$c^2$ at low $x_p$, where $x_p$
is the ratio between the meson and the beam energies, and no shift
at high $x_p$ ($x_p \sim 1$) \cite{act,laf}. OPAL also reported a
shift in the position of the $\rho^{\pm}$ peak from -10 to -30
MeV/$c^2$, which was consistent with the $\rho^0$ measurement
\cite{ack}. ALEPH reported the same shift on the position of
$\rho^0$ peak as observed by OPAL \cite{bus}. DELPHI fit the raw
invariant $\pi^+\pi^-$ mass distribution to the (BW$\times$PS +
BG) for $x_p >$ 0.05 and reported a $\rho^0$ mass of 757 $\pm$ 2
MeV/$c^2$ \cite{abr}, which is 7.5 standard deviations below the
$\rho^0$ mass in the vacuum (775.8 $\pm$ 0.5 MeV/$c^2$). As one
can see, similar to NA27, DELPHI assumed that the phase space was
described by the background function. Bose-Einstein correlations
were used to describe the shift on the position of $\rho^0$ peak.
However, high chaoticity parameters ($\lambda \sim 2.5$) were
needed \cite{act,laf,bus}. Previous measurements of the $\rho$
mass shift and possible explanations are discussed elsewhere
\cite{rho}. The masses of the $\rho^0$ and other short-lived
resonances in the vacuum are obtained only in exclusive reactions
and not in inclusive reactions where many particles are produced.

As previously mentioned \cite{rho}, one uncertainty in the
hadronic cocktail fit depicted in Fig.~\ref{fig:cocktail} is the
possible existence of correlations of unknown origin near the
$\rho^0$ mass. An example is correlations in the invariant mass
distribution from particles such as the $f_0(600)$ which are not
well established \cite{pdg}. The $\omega$ yield in the hadronic
cocktail fits may account for some of these contributions. In
order to evaluate the systematic uncertainty in the $\rho^0$ mass
due to poorly known contributions in the hadronic cocktail, the
$\rho^0$ mass was obtained by fitting the peak to the BW$\times$PS
function plus an exponential function representing these
contributions. Using this procedure, the $\rho^0$ mass is
systematically higher than the mass obtained from the hadronic
cocktail fit. This uncertainty is the main contribution to the
systematic uncertainties shown in Fig.~\ref{fig:massrho} and it
can be as large as $\sim$35 MeV/$c^2$ for low $p_T$. Other
contributions to the systematic errors shown in
Fig.~\ref{fig:massrho} result from uncertainty in the measurement
of particle momenta of $\sim$3 MeV/$c^2$ and from the hadronic
cocktail fits themselves of $\sim$10 MeV/$c^2$. The systematic
uncertainties are common to all $p_T$ bins and are correlated
between all centralities in the $d$+Au measurements.

\begin{figure}[tbp]
\centering
\includegraphics[width=0.45\textwidth]{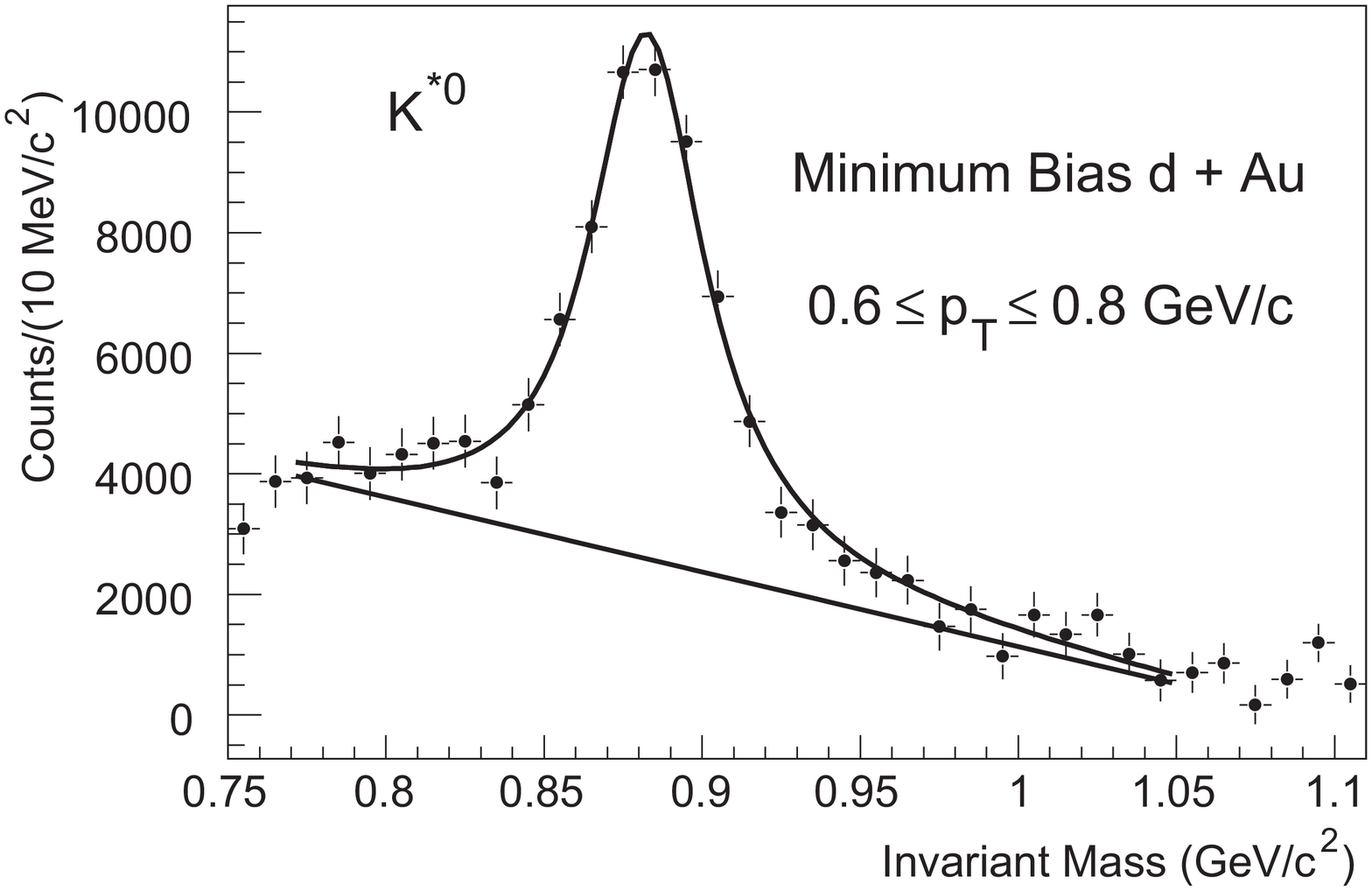}
\includegraphics[width=0.45\textwidth]{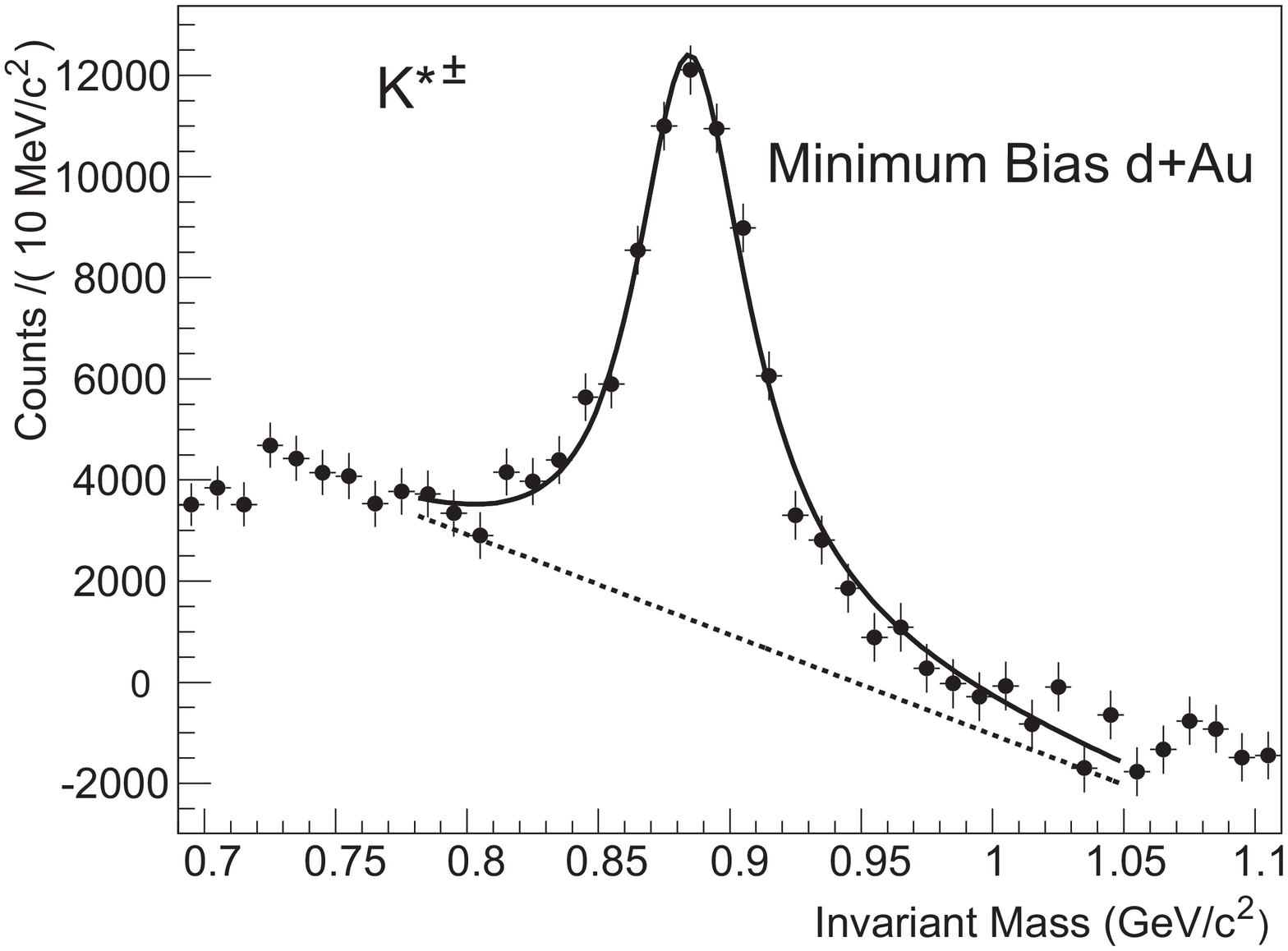}
\caption{The mixed-event background subtracted $K\pi$ raw
invariant mass distribution for a particular $K^{*0}$ $p_T$ bin
(upper panel) and the mixed-event background subtracted
$K_{S}^{0}\pi^{\pm}$ raw invariant mass distribution integrated
over the $K^{*\pm}$ $p_T$ (lower panel) at $|y| < 0.5$ for minimum
bias $d$+Au interactions. The dashed lines are the linear function
that describes the residual background.}\label{fig:kstar}
\end{figure}

Figure \ref{fig:kstar} depicts the mixed-event background
subtracted $K\pi$ and $K^0_S\pi^{\pm}$ invariant mass
distributions for minimum bias $d$+Au interactions at midrapidity
for a particular $p_T$ interval of the $K^{*0}$ $p_T$ and
integrated over the full measured $p_T$ range of the $K^{*\pm}$.
The signal to background ratio before subtraction is 1/50 for both
cases. The solid black line corresponds to the fit to the
relativistic $p$-wave Breit-Wigner function multiplied by the
phase space (equation \ref{bwstar}), with $T$ = 160 MeV, plus a
linear function that represents the residual background. This
comes predominantly from correlated $K\pi$ pairs and correlated
but misidentified pairs. A detailed study has been presented
previously \cite{kstar200}.

\begin{figure}[tbp]
\centering
\includegraphics[width=0.45\textwidth]{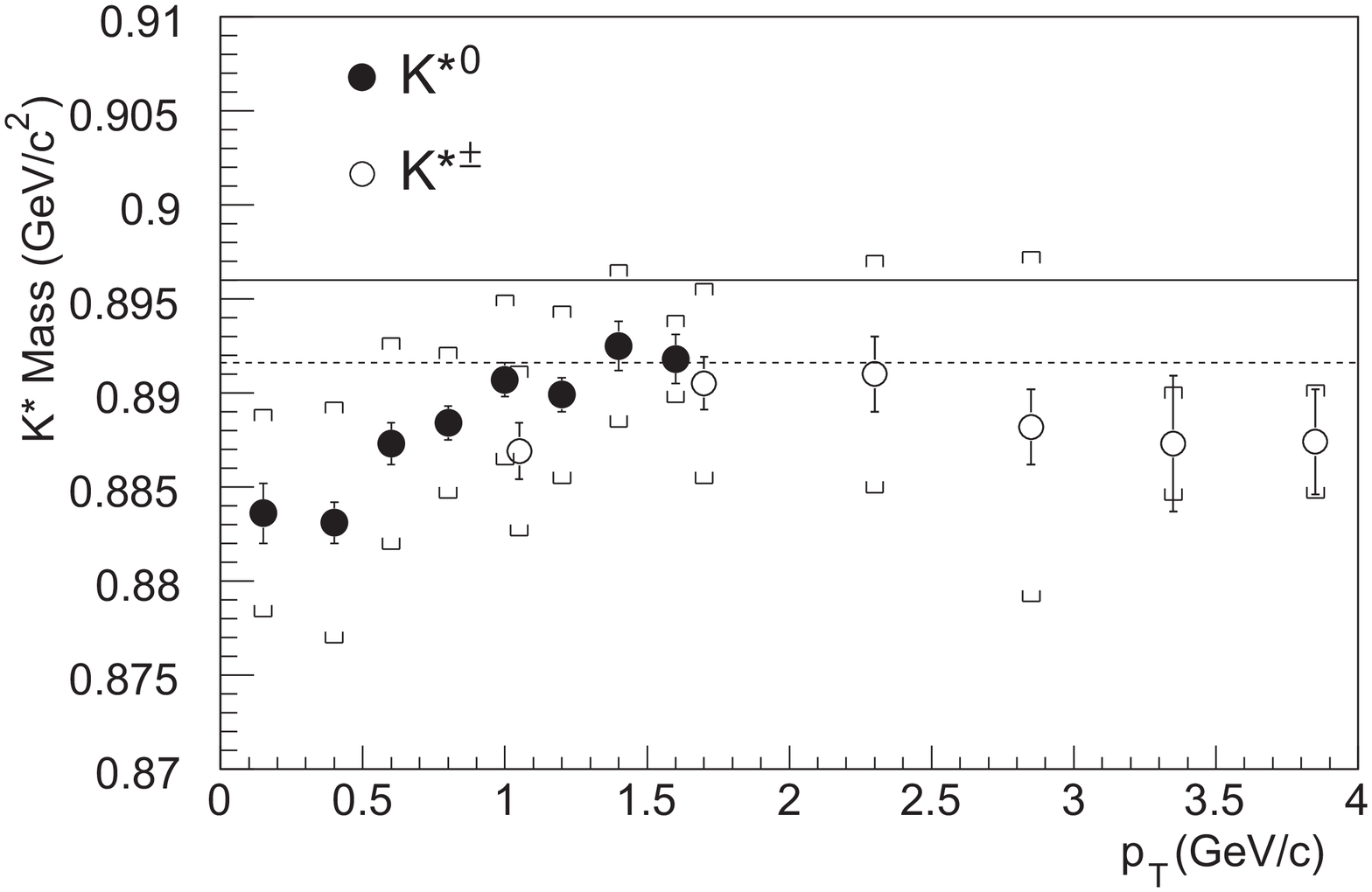}
\includegraphics[width=0.45\textwidth]{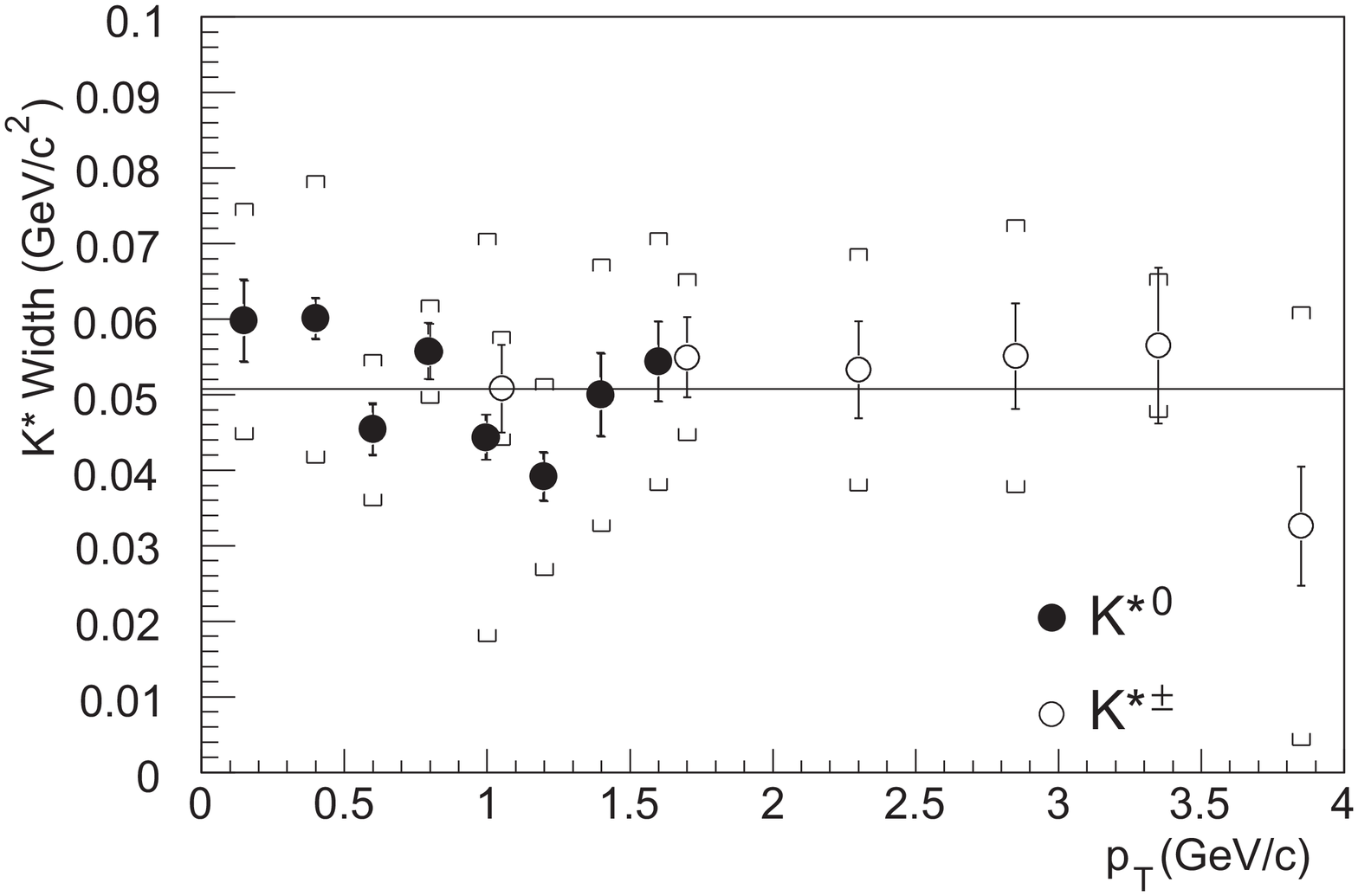}
\caption{The $K^{*}$ mass (upper panel) and width (lower panel) as
a function of $p_T$ at $|y| < 0.5$ for minimum bias $d$+Au
collisions. In the upper panel, the solid line is the PDG $K^{*0}$
mass (0.8961 GeV/$c^2$) \cite{pdg}. The dashed line is the PDG
$K^{*\pm}$ mass (0.8917 GeV/$c^2$) \cite{pdg}. In the lower panel,
the solid line is the  $K^{*0}$ and $K^{*\pm}$ widths (0.0507
GeV/$c^2$) \cite{pdg}. The brackets indicate the systematic
uncertainties.}\label{fig:kstarmasswidth}
\end{figure}

The $K^*$ masses and widths at $|y| < 0.5$ for minimum bias $d$+Au
interactions as a function of $p_T$ are depicted in Fig.
\ref{fig:kstarmasswidth}. Both mass and width were obtained by
correcting the $K^*$ distribution for detector acceptance and
efficiency that was determined from a detailed simulation of the
TPC response using GEANT \cite{detectorres}. The $K^{*0}$ mass
increases as a function of $p_T$ and at low $p_T$ ($p_T < $1.1
GeV/$c$) the mass is significantly smaller than previously
reported values \cite{pdg}. A similar mass shift was observed in
minimum bias $p+p$ collisions at $\sqrt{s_{NN}}$ = 200 GeV
\cite{kstar200} and the possible explanations are the same as
described for the $\rho^0$ meson. Even though a $K^{*0}$ mass
shift in $d$+Au collisions has not been observed before, it is
important to note that previous measurements were mainly
interested in extracting the resonance cross-section \cite{rhopp}.
In addition, we observe a mass shift at low $p_T$ of $\sim$ 10
MeV/$c^2$, while previous analyses only presented the $K^{*0}$
mass integrated in $p_T$, $x_F$, or $x_p$. The $K^{*\pm}$ mass is
in agreement with previous values within errors \cite{pdg}.
However, this could be due to the limited $p_T$ range covered.
There is no significant difference between the measured $K^*$
width and the previous values \cite{pdg}.

The main contributions to the systematic uncertainties on the
$K^{*}$ mass and width were evaluated as a function of $p_T$ using
a different residual background function (second order
polynomial), different fitting functions to the $K^{*}$ invariant
mass (non-relativistic BW, relativistic BW without phase-space
factor), and different slope parameters in the BW$\times$PS
function (140 MeV and 180 MeV). In addition, the mass and the
width were obtained separately for $K^{*0}$, $\overline{K}^{*0}$,
$K^{*+}$, and $K^{*-}$. The systematic uncertainty due to detector
effects was also accounted for. The systematic uncertainty can be
as large as $\sim$6.5 (9.5) MeV/$c^2$ and $\sim$25 (30) MeV/$c^2$
for the $K^{*0} (K^{*\pm})$ mass and width, respectively.

\begin{figure}[tbp]
\centering
\includegraphics[width=0.45\textwidth]{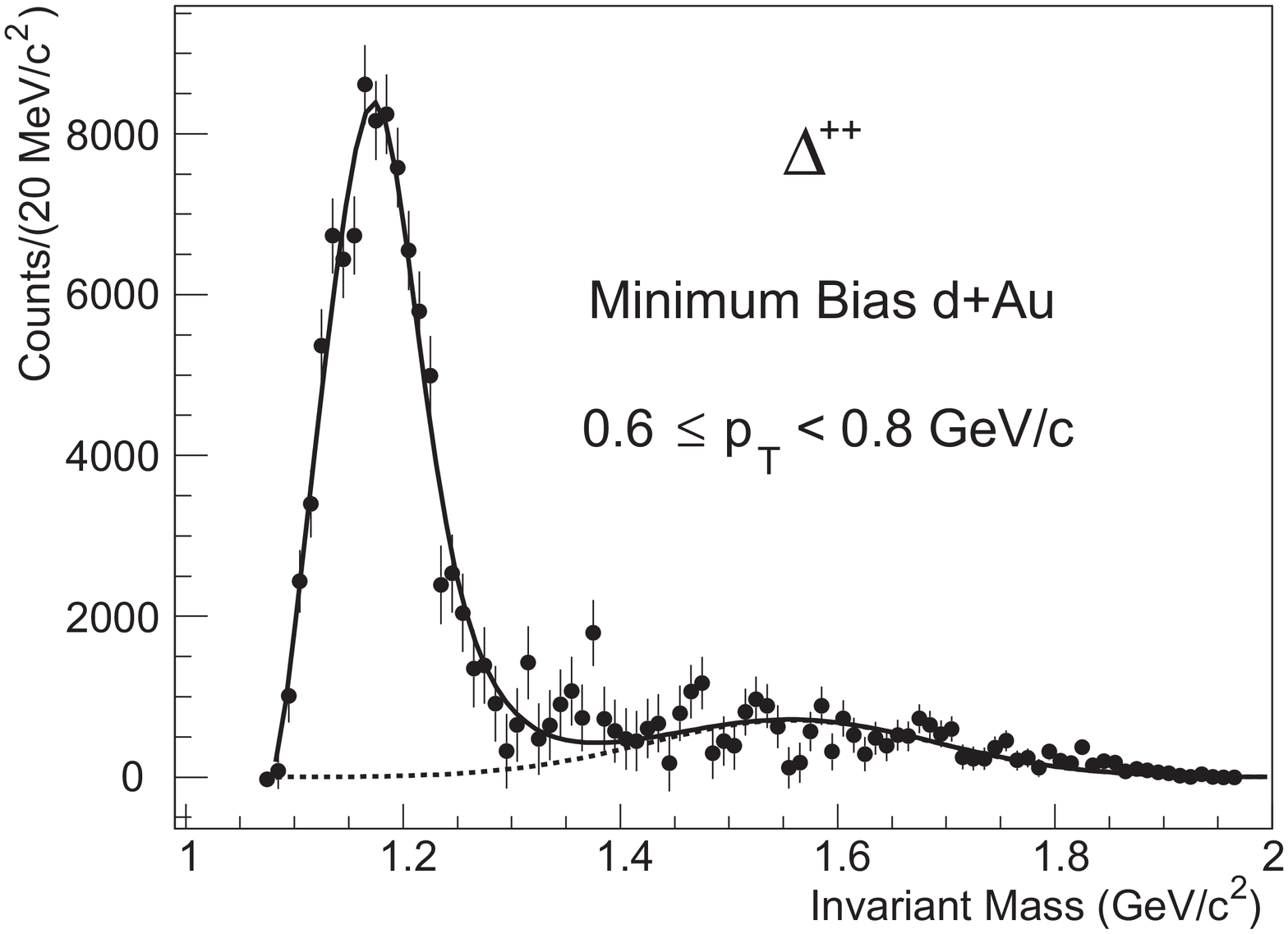}
\includegraphics[width=0.45\textwidth]{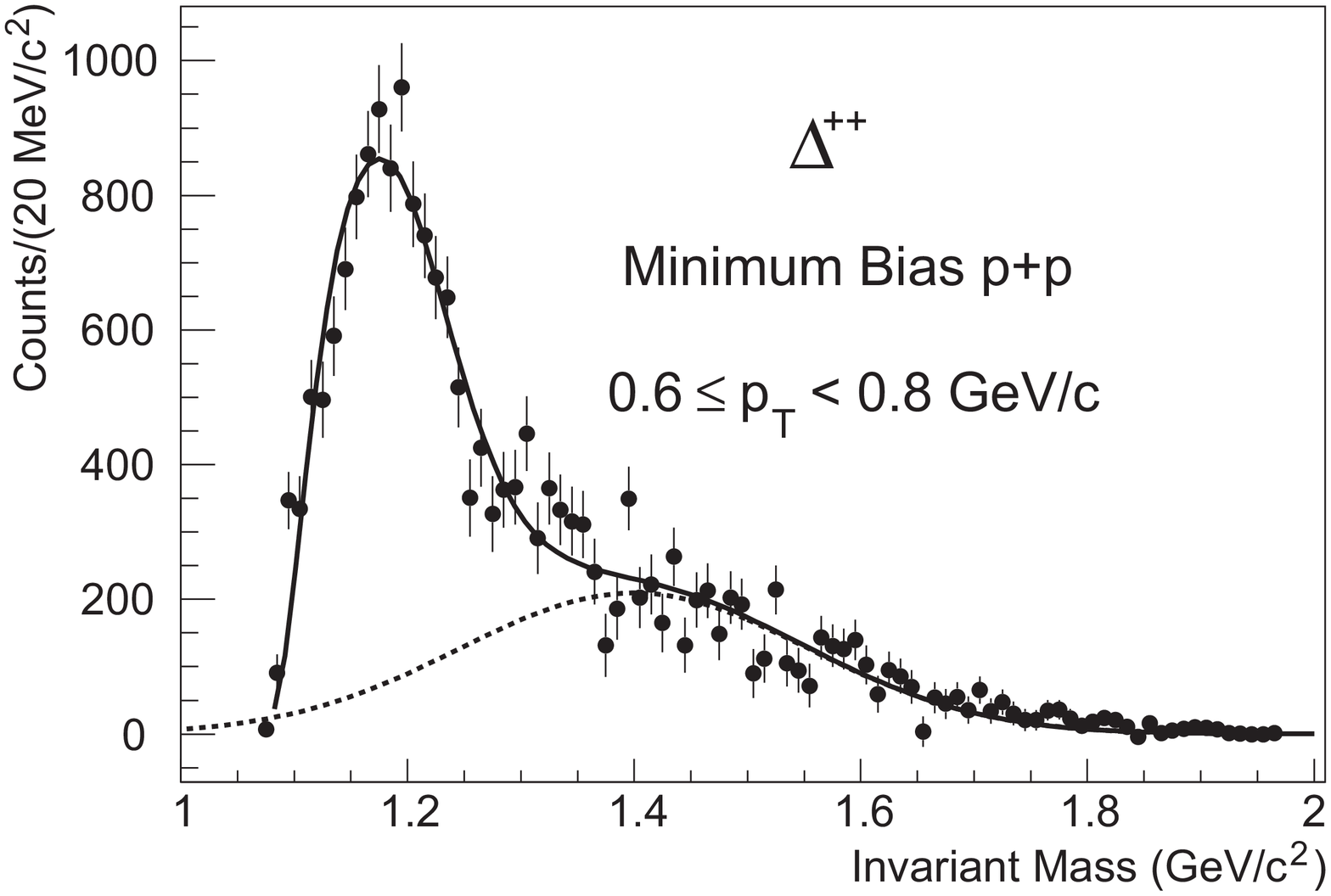}
\caption{The mixed-event background subtracted $p\pi$ raw
invariant mass distribution for a particular $p_T$ bin at $|y| <
0.5$ for minimum bias $d$+Au collisions (upper panel) and for
minimum bias $p+p$ collisions (lower panel). The dashed lines are
the linear function that describes the residual background.}
\label{fig:delta}
\end{figure}

The $p\pi$ raw invariant mass distributions after the mixed-event
background subtraction for minimum bias $d$+Au and $p+p$
interactions at midrapidity for a particular $p_T$ bin are shown
in Fig. \ref{fig:delta}. Before background subtractions, the
signal to background ratios are 1/50 and 1/30 for minimum bias
$d$+Au and $p+p$ interactions, respectively. The solid black line
corresponds to the fit to a relativistic $p$-wave Breit-Wigner
function multiplied by the phase space, with $T$ = 160 MeV, plus a
Gaussian function that represents the residual background
indicated by a dashed line. In this case, the normalization factor
used to subtract the combinatorial background was changed until
the best $\chi^2/ndf$ was achieved. Similar to the $\rho^0$
analysis \cite{rho}, the uncorrected yield of the $\Delta^{++}$
was a free parameter in the fit and the $\Delta^{++}$ distribution
was corrected for the detector acceptance and efficiency
determined from a detailed simulation of the TPC response using
GEANT \cite{detectorres}. The relativistic $p$-wave Breit-Wigner
function multiplied by the phase space is the same as equation
\ref{bwstar}. However, in the case of the $\Delta^{++}$, the mass
dependent width is given by:
\begin{equation}
\Gamma = \frac{\Gamma_0M_0}{M_{p\pi}} \times \frac{k(M_{p\pi})^3
F(\Lambda_{\pi},k(M_{p\pi}))^2}{k(M_0)^3
F(\Lambda_{\pi},k(M_0))^2}
\end{equation}
where $F(\Lambda_{\pi},k_{CM})$ is the form factor used to fit the
$\pi - N$ scattering phase-shift with $\Lambda_{\pi} = 290$
MeV/$c^2$ \cite{deltaformfactor}, and
\begin{equation}
k(M_{p\pi})^2 = \frac{(M_{p\pi}^2 - m_p^2 -
m_{\pi}^2)^2-4m_p^2m_{\pi}^2}{4M_{p\pi}^2}.
\end{equation}

\begin{figure}[tbp]
\centering
\includegraphics[width=0.45\textwidth]{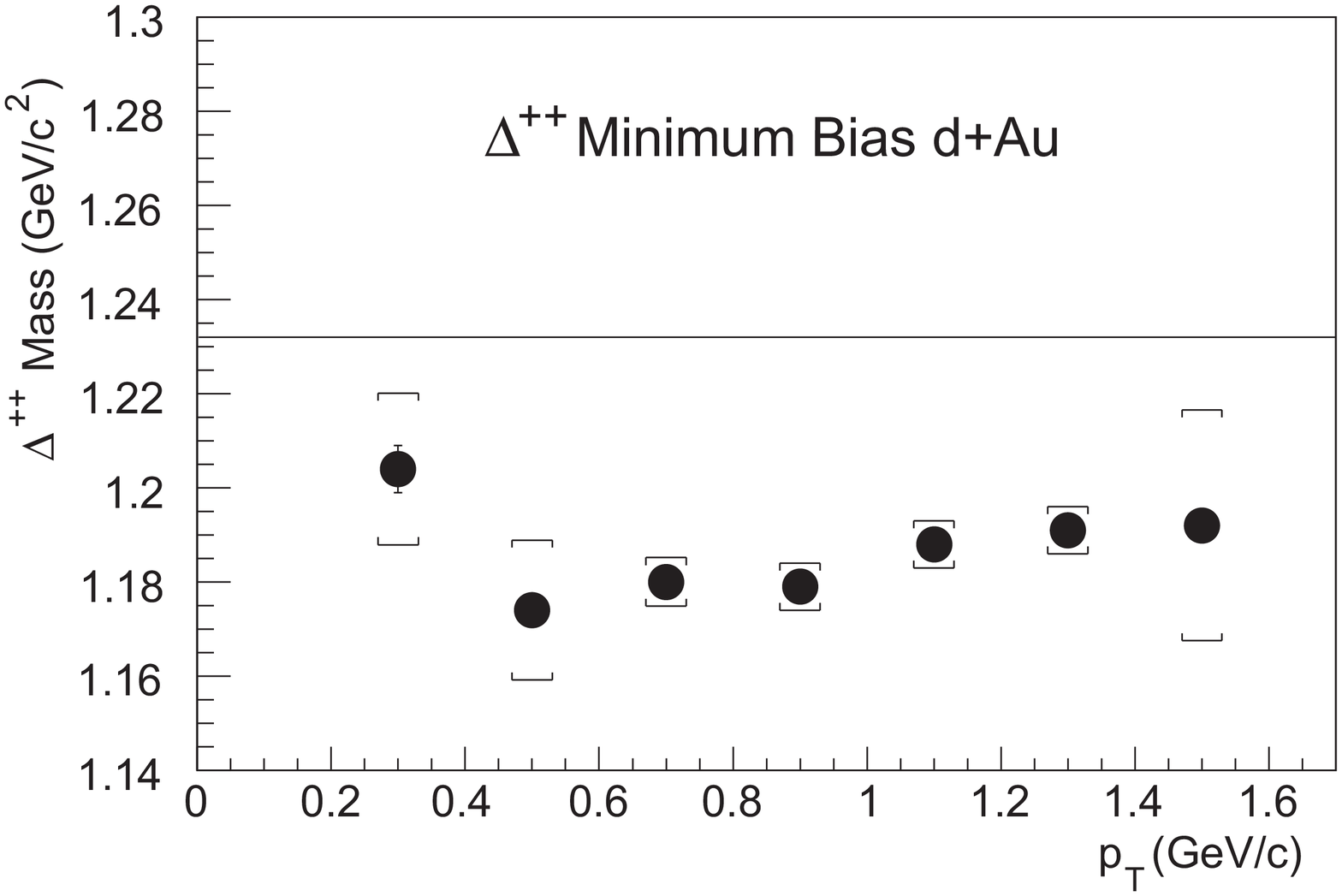}
\includegraphics[width=0.45\textwidth]{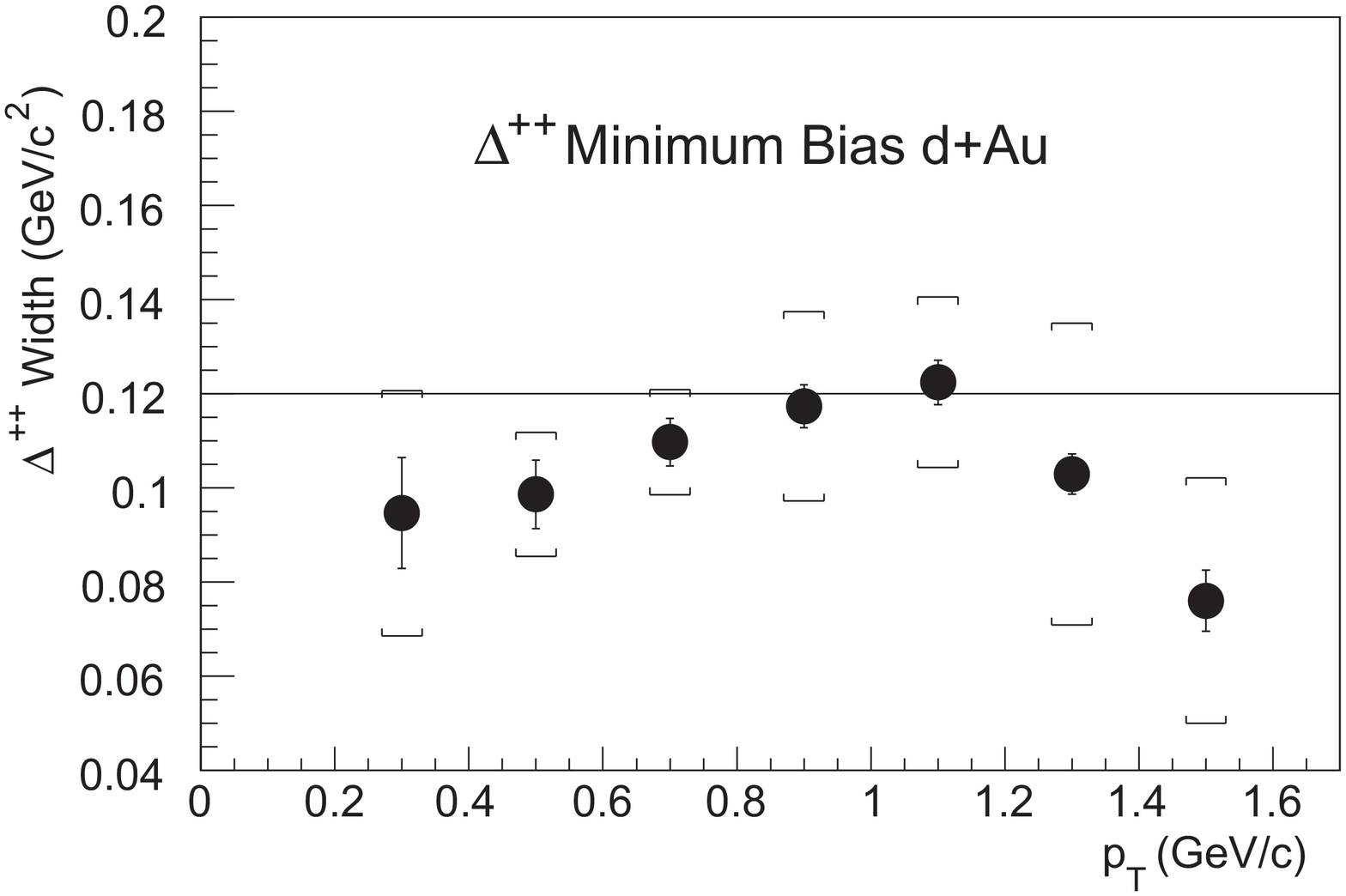}
\caption{The $\Delta^{++}$ mass (upper panel) and width (lower
panel) as a function of $p_T$ at $|y| < 0.5$ for minimum bias
$d$+Au collisions. The solid lines are the PDG $\Delta^{++}$ mass
(1.232 GeV/$c^2$) and width (0.120 GeV/$c^2$) \cite{pdg}. The
brackets indicate the systematic
uncertainties.}\label{fig:deltamasswidth}
\end{figure}

The $\Delta^{++}$ mass and width at $|y| < 0.5$ for minimum bias
$d$+Au interactions as a function of $p_T$ are depicted in Fig.
\ref{fig:deltamasswidth}. The $\Delta^{++}$ mass is significantly
smaller than the values previously reported, though the width is
in agreement within errors \cite{pdg}. Possible explanations for a
$\Delta^{++}$ mass shift are the same as for the $\rho^0$
\cite{rho}.

\begin{figure}[tbp]
\centering
\includegraphics[width=0.45\textwidth]{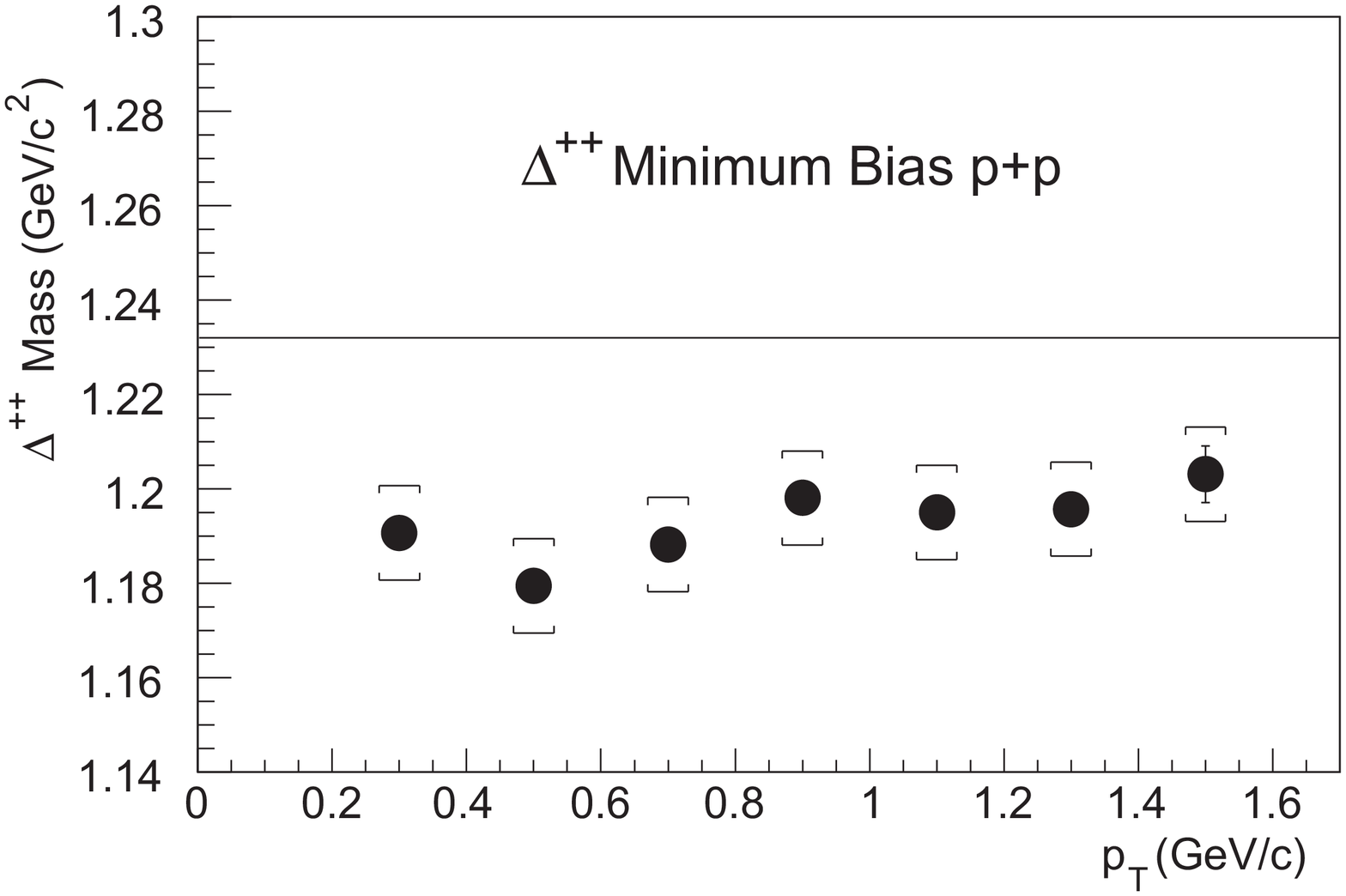}
\includegraphics[width=0.45\textwidth]{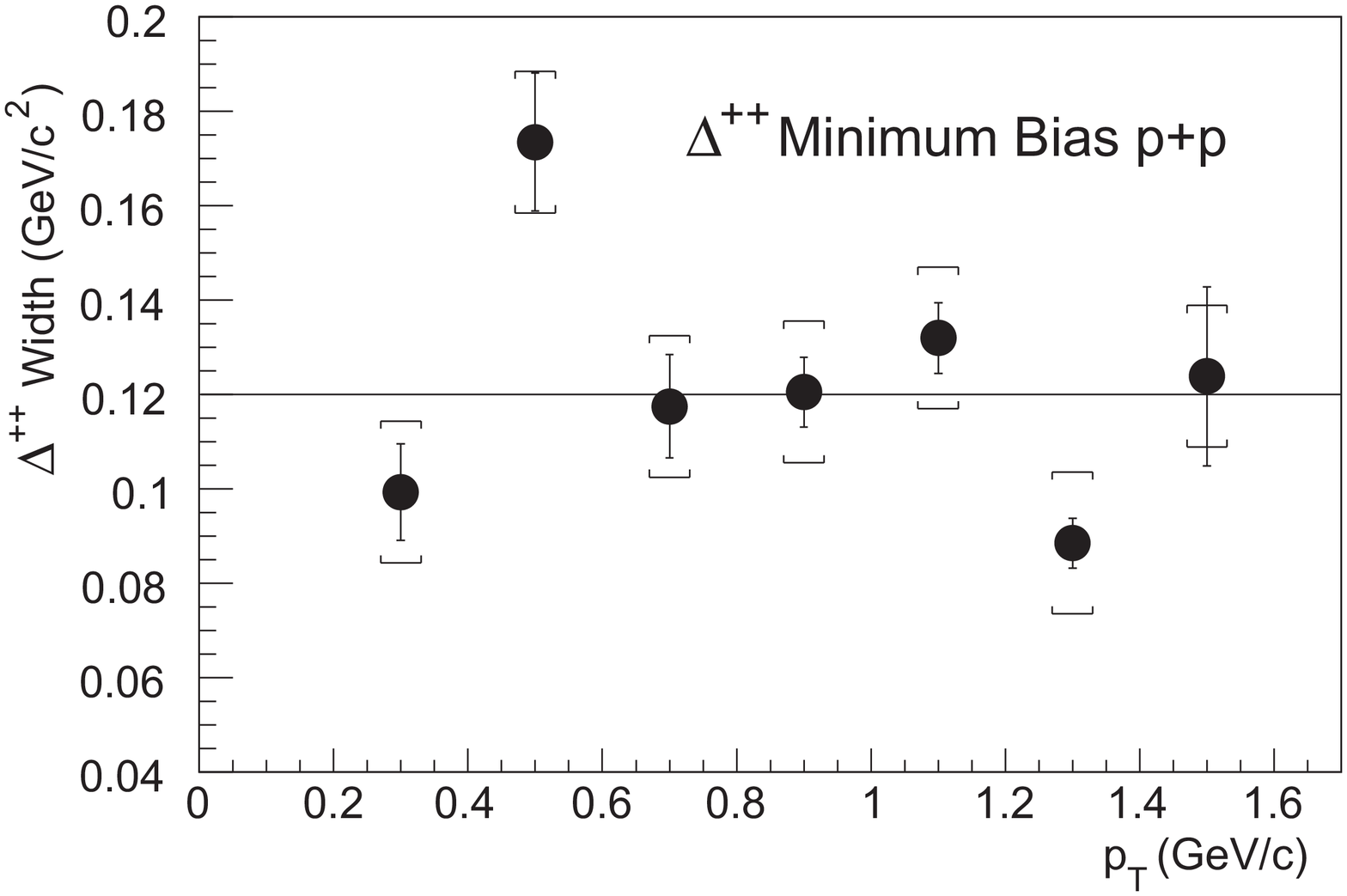}
\caption{The $\Delta^{++}$ mass (upper panel) and width (lower
panel) as a function of $p_T$ at midrapidity for minimum bias
$p+p$ interactions. The solid lines are the PDG $\Delta^{++}$ mass
(1.232 GeV/$c^2$) and width (0.120 GeV/$c^2$) \cite{pdg}. The
brackets indicate the systematic
uncertainties.}\label{fig:deltamasswidthpp}
\end{figure}

The $\Delta^{++}$ mass and width at midrapidity for minimum bias
$p+p$ collisions as a function of $p_T$ are shown in Fig.
\ref{fig:deltamasswidthpp}. The analysis procedure in minimum bias
$p+p$ is the same as in $d$+Au collisions. Similarly to the $d$+Au
measurement, the $\Delta^{++}$ mass is significantly smaller than
the values in \cite{pdg} and the same possible explanations apply.
The $\Delta^{++}$ width is in agreement with previous values
 within errors \cite{pdg}.

In the case of the $\Delta^{++}$ mass and width, the main
contributions to the systematic uncertainties were calculated as a
function of $p_{T}$ by using different residual background
functions (first and second-order polynomial) and different slope
parameters in the BW$\times$PS function. The mass and the width
were also obtained separately for $\Delta^{++}$ and
$\overline{\Delta}^{--}$. The contribution due to the uncertainty
in the measurement of particle momenta is $\sim$5 MeV/$c^2$. The
systematic uncertainty can be as large as $\sim$20 MeV/$c^2$ and
$\sim$30 MeV/$c^2$ for the $\Delta^{++}$ mass and width,
respectively. In $p+p$ collisions, the systematic uncertainty on
the mass and the width was evaluated similarly to the measurement
in $d$+Au collisions. The systematic uncertainty is $\sim$10
MeV/$c^2$ and $\sim$15 MeV/$c^2$ for the $\Delta^{++}$ mass and
width, respectively.

\begin{figure}[tbp]
\centering
\includegraphics[width=0.45\textwidth]{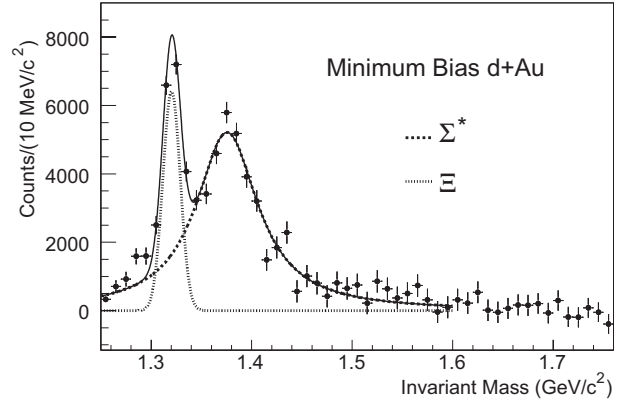}
\caption{The mixed-event background subtracted $\Lambda\pi$ raw
invariant mass distribution integrated over the $\Sigma^*$ $p_T$
at $|y| < 0.75$ for minimum bias $d$+Au collisions.}
\label{fig:sigma}
\end{figure}

\begin{figure}[tbp]
\centering
\includegraphics[width=0.45\textwidth]{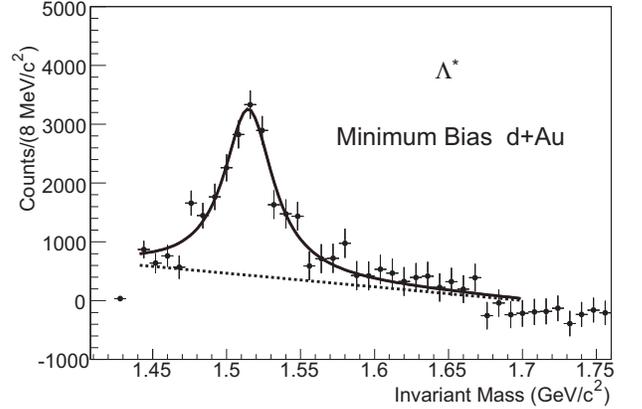}
\caption{The mixed-event background subtracted $pK$ raw invariant
mass distribution integrated over the $\Lambda^*$ $p_T$ at $|y| <
0.5$ for minimum bias $d$+Au collisions. The dashed line is the
linear function that describes the residual background.}
\label{fig:lambda}
\end{figure}

The mid-rapidity and $p_T$ integrated $\Lambda\pi$ ($\Sigma^*$)
and $pK$ ($\Lambda^*$) raw invariant mass distributions, after the
mixed-event background subtraction, from minimum bias $d$+Au
collisions are depicted in Fig. \ref{fig:sigma} and Fig.
\ref{fig:lambda}, respectively. The signal to background ratio is
1/14 for the $\Sigma^*$ and 1/24 for the $\Lambda^*$ before
mixed-event background subtraction. Since the $\Xi^-$ and the
$\overline{\Xi}^+$ have the same final state as the $\Sigma^{*-}$
and $\overline{\Sigma}^{*+}$, the $\Lambda\pi$ invariant mass
distribution is fitted to a Gaussian combined with a
non-relativistic Breit-Wigner function (SBW)
\begin{equation}
\mathrm{SBW} = \frac{\Gamma_0}{(M_{\Lambda\pi} - M_0)^2 +
\Gamma^2/4}
\end{equation}
In the case of the $\Lambda^*$, the signal is fitted to a
non-relativistic Breit-Wigner function combined with a linear
function that describes the residual background \cite{baryon}.

\begin{figure}[tbp]
\centering
\includegraphics[width=0.45\textwidth]{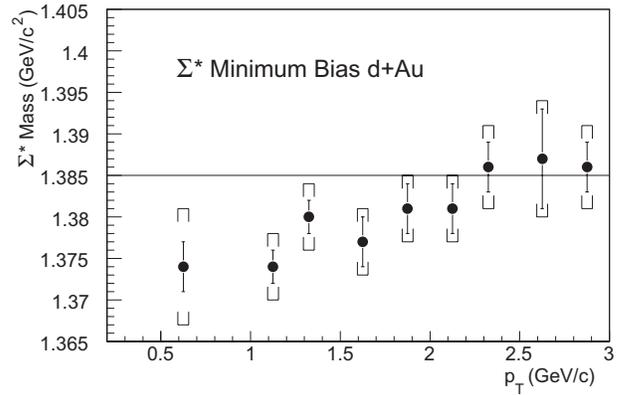}
\caption{The $\Sigma^*$ mass as a function of $p_T$ at $|y| <
0.75$ for minimum bias $d$+Au collisions. The solid line is the
PDG $\Sigma^*$ mass average between the masses of $\Sigma^{*+}$ and
$\Sigma^{*-}$ (1.3850 GeV/$c^2$) \cite{pdg}. The brackets indicate
the systematic uncertainties.} \label{fig:sigmamass}
\end{figure}

\begin{figure}[tbp]
\centering
\includegraphics[width=0.45\textwidth]{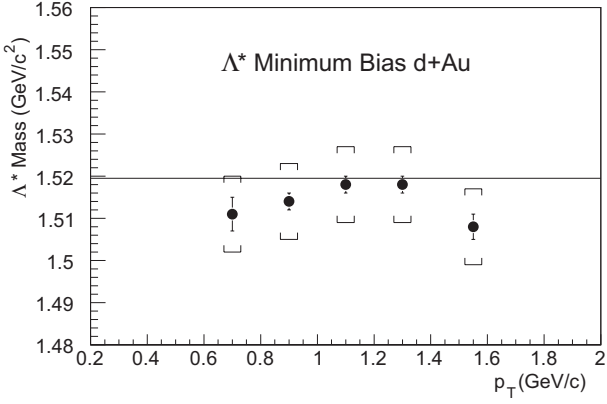}
\caption{The $\Lambda^*$ mass as a function of $p_T$ at $|y| <
0.5$ for minimum bias $d$+Au collisions. The solid line is the PDG
$\Lambda^*$ mass (1.5195 GeV/$c^2$) \cite{pdg}. The brackets
indicate the systematic uncertainties.} \label{fig:lambdamass}
\end{figure}

\begin{figure}[tbp]
\centering
\includegraphics[width=0.45\textwidth]{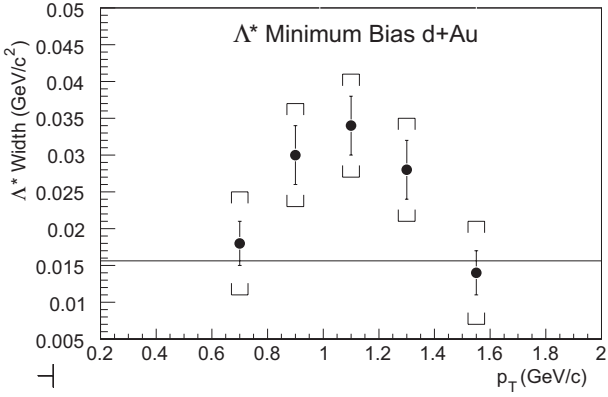}
\caption{The $\Lambda^*$ width as a function of $p_T$ at $|y| <
0.5$ for minimum bias $d$+Au collisions. The solid line is the PDG
$\Lambda^*$ width (0.0156 GeV/$c^2$) \cite{pdg}. The brackets
indicate the systematic uncertainties. Our simulations, including
both the detector resolution and kinetic cuts, show a width of
0.022 GeV/$c^2$.} \label{fig:lambdawidth}
\end{figure}

The fit parameters corresponding to the $\Sigma^{*}$ mass and
width in the integrated $p_{T}$ interval (0.25 $< p_{T} <$ 3.5
GeV/$c$) are $1.376 \pm 0.002 \pm 0.007$ GeV/$c^{2}$ and $48
\pm 2 \pm 10$ MeV/$c^{2}$, respectively. Both the measured width
and the mass, within their uncertainty, are in agreement with the
PDG values of the $\Sigma^{*}$ \cite{pdg}. The systematic errors
include the uncertainty due to choice of bin size, the
normalization of the mixed event background, the variations in the
fit range and the selections of event and tracks. It is possible
to further study the $p_{T}$ dependence of the $\Sigma^{*}$ mass
when the width is fixed to the PDG value ($37.6 \pm 1.1$
MeV/$c^{2}$) \cite{pdg} and the mass is a free parameter in the
Breit-Wigner function. Figure \ref{fig:sigmamass} shows the
$p_{T}$ dependence for the $\Sigma^{*}$ mass from the fit
function. There is a small difference in the mass for low $p_T$
$\Sigma^*$ compared to the PDG value.

The results for the $\Lambda^*$ mass and width are shown in Figs.
\ref{fig:lambdamass} and \ref{fig:lambdawidth}, respectively. The
$\Lambda^*$ mass obtained from the data is 1515.0 $\pm$ 1.2$\pm$ 3
MeV/$c^2$, consistent with the $\Lambda^*$ natural mass of 1519.5
$\pm$ 1.0 MeV/$c^2$ \cite{pdg} within errors. The width of the
$p_T$ integrated spectrum is 40 $\pm$ 5 $\pm$ 10 MeV/$c^2$ which
includes the intrinsic resolution of the detector
\cite{detectorres} of 6 MeV and the momentum dependent mass shifts
in the data, which are in the statistical and systematical limits.
The measured width in each momentum bin is consistent with folding
the $\Lambda^*$ natural width of 15.6 $\pm$ 1.0 MeV/$c^2$
\cite{pdg} with the detector resolution. The systematic
uncertainties are due to the residual background, the range used
for the normalization and for the fit to the signal, and different
bin widths.

In $d$+Au collisions, we observe modifications of the mass and decay width
of short-lived resonances that might be due to dynamical interactions with the surrounding
matter, interference, phase space, and Bose-Einstein correlations \cite{rho}.

\subsection{Spectra}

In $p+p$ collisions at RHIC, a shape
difference in the $p_T$ spectra of mesons and
baryons for non-resonant particles in the interval $2 < p_T < 6 $ GeV/$c$
at $\sqrt{s_{_{NN}}}$ = 200 GeV was observed \cite{starpp}. In order to verify
if such effect is observed in $d$+Au
collisions for resonances, their spectra are studied.

The uncorrected yields obtained in each $p_T$ bins were corrected
for the detector acceptance and efficiency determined from a
detailed simulation of the TPC response using GEANT
\cite{detectorres}. The yields were also corrected for the
corresponding branching ratios listed in Table \ref{tab:BR}, to
account for the fact that we only measure certain decay modes.

\begin{figure}[tbp]
\centering
\includegraphics[width=0.45\textwidth]{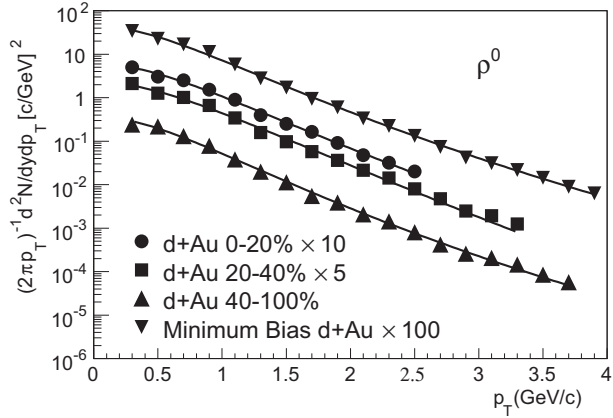}
\caption{The $\rho^0$ invariant yields as a function of $p_T$ at
$|y|<$ 0.5 for minimum bias $d$+Au interactions and 0-20$\%$,
20-40$\%$, and 40-100$\%$ of the total $d$+Au cross-section. The
lines are fits to a Levy function (equation \ref{levy}). The
errors are statistical only and smaller than the
symbols.}\label{fig:rhodndy}
\end{figure}

\begin{figure}[tbp]
\centering
\includegraphics[width=0.45\textwidth]{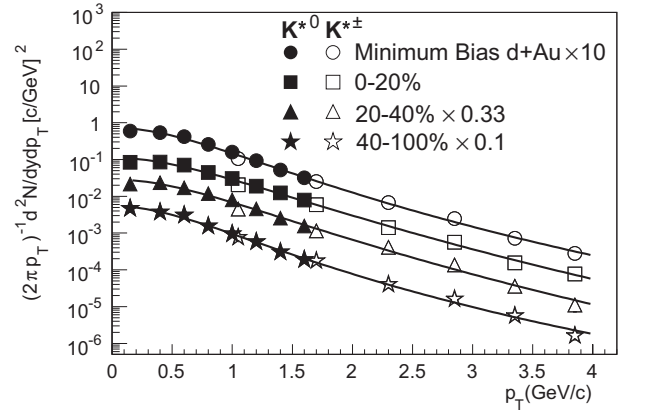}
\caption{The ($K^{*}+\overline{K}^{*}$)/2 together with the
($K^{*+}+{K^{*-}}$)/2 invariant yields as a function of $p_T$ at
$|y|<$ 0.5 for minimum bias $d$+Au collisions and three different
centralities. The lines are fits to a Levy function (equation
\ref{levy}). The errors are statistical only and smaller than the
symbols.}\label{fig:kstardndy}
\end{figure}

The $\rho^0$ and the ($K^{*}+\overline{K^{*}}$)/2 corrected
invariant yields [$d^2N/(2\pi p_Tdp_Tdy)$] at $|y| < $ 0.5 as a
function of $p_T$ for minimum bias $d$+Au interactions are shown
in Fig. \ref{fig:rhodndy} and Fig. \ref{fig:kstardndy},
respectively. A Levy function \cite{kstar200}
\begin{eqnarray}
\frac{1}{2\pi p_T} \frac{d^2N}{dydp_T} = \frac{dN}{dy} \times
\frac{(n-1)(n-2)}{2\pi nT(nT+m_0(n-2))} \times \nonumber
\\
(1+\frac{\sqrt{p_T^2+m_0^2}-m_0}{nT})^{-n},
\label{levy}
\end{eqnarray}
was used to extract the $\rho^0$ and $K^*$ yields per unit of
rapidity around midrapidity. In the limit of low $p_T$, the Levy
function is an exponential function and a power law in the limit
of high $p_T$.

\begin{figure}[tbp] \centering
\includegraphics[width=0.45\textwidth]{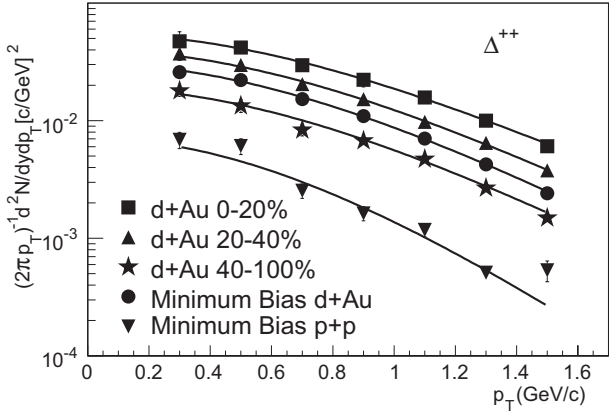}
\caption{The ($\Delta^{++}+\overline{\Delta}^{--})/2$ invariant
yields as a function of $p_T$ at $|y|<$ 0.5 for minimum bias
$p+p$, $d$+Au interactions and 0-20$\%$, 20-40$\%$, and 40-100$\%$
of the total $d$+Au cross-section. The lines are fits to an
exponential function (equation \ref{exp}). The errors are
statistical only and smaller than the symbols for the spectra
measured in $d$+Au. In the $p+p$ measurement, the errors shown
also include the systematic uncertainties.}\label{fig:deltadndy}
\end{figure}

\begin{figure}[tbp]
\centering
\includegraphics[width=0.45\textwidth]{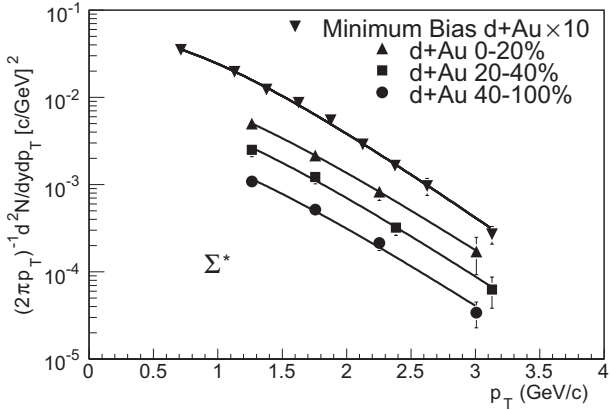}
\caption{The $\Sigma^*$ invariant yields as a function of $p_T$ at
$|y|<$ 0.75 for minimum bias $d$+Au collisions and three different
centralities. The lines are fits to an exponential function
(equation \ref{exp}).}\label{fig:sigmadndy}
\end{figure}

\begin{figure}[tbp]
\centering
\includegraphics[width=0.45\textwidth]{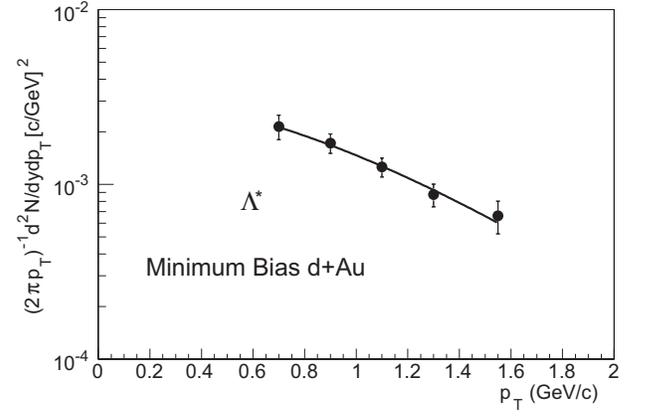}
\caption{The $\Lambda^*$ invariant yields as a function of $p_T$
at $|y|<$ 0.5 for minimum bias $d$+Au collisions. The line is a
fit to an exponential function (equation
\ref{exp}).}\label{fig:lambdadndy}
\end{figure}

The ($\Delta^{++}$+$\overline\Delta^{--}$)/2,
($\Sigma^{*-}$+$\overline\Sigma^{*+}$)/2, and
($\Lambda^*$+$\overline\Lambda^*$)/2 corrected invariant yields
 at $|y| < $ 0.5 as a function of $p_T$
are shown in Fig. \ref{fig:deltadndy}, Fig. \ref{fig:sigmadndy},
and Fig. \ref{fig:lambdadndy}, respectively. Figure
\ref{fig:deltadndy} also depicts the
($\Delta^{++}$+$\overline\Delta^{--}$)/2 corrected invariant yield
for minimum bias $p+p$. Since the $p_T$ region is limited to low
$p_T$, we use an exponential function \cite{kstar200}
\begin{eqnarray}
\frac{1}{2\pi m_T} \frac{d^2N}{dydm_T} = \frac{dN}{dy} \times
\frac{1}{2\pi T (m_0+T)} \times \nonumber
\\
\exp(\frac{-(m_T-m_0)}{T}),
\label{exp}
\end{eqnarray}
to extract the $\Delta^{++}$, $\Sigma^*$, and $\Lambda^*$ yields
per unit of rapidity around midrapidity. Due to limited
statistics, only the $\Lambda^*$ yield in minimum bias $d$+Au
collisions was measured.

\begin{table*}
\caption{\label{tab:rhokstardndy}The $\rho^0$ and ($K^* +
\overline{K}^*$)/2 $dN/dy$, $T$, and $n$ at $|y|<$ 0.5 measured in
minimum bias $d$+Au collisions and 0-20$\%$, 20-40$\%$, and
40-100$\%$ of the total $d$+Au cross-section. The first error is
statistical; the second is systematic.}
\begin{ruledtabular}
\begin{tabular}{ccccc}
Resonance & Centrality & $dN/dy$ & $T$ (MeV) & $n$ \\
\hline
    $\rho^0$     & Minimum Bias $d$+Au & 0.812 $\pm$ 0.004 $\pm$ 0.085 & 231.7 $\pm$ 1.6 $\pm$ 35 & 11.1 $\pm$ 0.2 \\
                 & 0-20$\%$            & 1.169 $\pm$ 0.014 $\pm$ 0.17 & 245 $\pm$ 6 $\pm$ 52 & 13.6 $\pm$ 1.7  \\
                 & 20-40$\%$           & 0.958 $\pm$ 0.011 $\pm$ 0.14 & 230 $\pm$ 4 $\pm$ 44 & 11.4 $\pm$ 0.6 \\
                 & 40-100$\%$          & 0.607 $\pm$ 0.005 $\pm$ 0.13 & 212 $\pm$ 3 $\pm$ 36 & 10.7 $\pm$ 0.4 \\
    $K^*$        & Minimum Bias $d$+Au & 0.161 $\pm$ 0.002 $\pm$ 0.027 & 286 $\pm$ 7 $\pm$ 44 & 10.4 $\pm$ 0.1 \\
                 & 0-20$\%$            & 0.294 $\pm$ 0.009 $\pm$ 0.051 & 316 $\pm$ 22 $\pm$ 53 & 12.8 $\pm$ 0.4 \\
                 & 20-40$\%$           & 0.204 $\pm$ 0.005 $\pm$ 0.037 & 306 $\pm$ 17 $\pm$ 50 & 12.5 $\pm$ 0.3 \\
                 & 40-100$\%$          & 0.108 $\pm$ 0.002 $\pm$ 0.018 & 232 $\pm$ 7 $\pm$ 39 & 7.3 $\pm$ 0.6 \\
\end{tabular}
\end{ruledtabular}
\end{table*}

\begin{table*}
\caption{\label{tab:deltasigmalambdadndy}The $(\Delta^{++} +
\overline{\Delta}^{--})/2$, $(\Sigma^* + \overline{\Sigma}^*)/2$,
and $(\Lambda^* + \overline{\Lambda}^*)/2$ $dN/dy$ and $T$ at
$|y|<$ 0.5 measured in minimum bias $d$+Au collisions and
0-20$\%$, 20-40$\%$, and 40-100$\%$ of the total $d$+Au
cross-section. In the case of the $\Delta^{++}$, the results from
the measurements in minimum bias $p+p$ are also shown. The first error is
statistical; the second is systematic.}
\begin{ruledtabular}
\begin{tabular}{cccc}
Resonance & Centrality & $dN/dy$ & $T$ (MeV) \\
\hline
    $\Delta^{++}$ & Minimum Bias $d$+Au & 0.0823 $\pm$ 0.0012 $\pm$ 0.0099 & 284 $\pm$ 7 $\pm$ 45\\
                  & 0-20$\%$            & 0.177 $\pm$ 0.005 $\pm$ 0.021 & 328 $\pm$ 17 $\pm$ 52\\
                  & 20-40$\%$           & 0.116 $\pm$ 0.003 $\pm$ 0.014 & 303 $\pm$ 14$ \pm$ 48\\
                  & 40-100$\%$          & 0.0529 $\pm$ 0.0008 $\pm$ 0.0063 & 290 $\pm$ 9 $\pm$ 46\\
                  & Minimum Bias $p+p$  & 0.0139 $\pm$ 0.0008 $\pm$ 0.0050 & 216 $\pm$ 13 $\pm$ 86 \\
    $\Sigma^*$    & Minimum Bias $d$+Au & 0.0319 $\pm$ 0.0011 $\pm$ 0.0041 & 387 $\pm$ 11 $\pm$ 28 \\
                  & 0-20$\%$            & 0.068 $\pm$ 0.006 $\pm$ 0.011 & 473 $\pm$ 39 $\pm$ 40 \\
                  & 20-40$\%$           & 0.040 $\pm$ 0.004 $\pm$ 0.007 & 420 $\pm$ 36 $\pm$ 40 \\
                  & 40-100$\%$          & 0.018 $\pm$ 0.002 $\pm$ 0.005 & 428 $\pm$ 36 $\pm$ 40 \\
    $\Lambda^*$   & Minimum Bias $d$+Au & 0.0149 $\pm$ 0.0014 $\pm$ 0.0022 & 392 $\pm$ 75 $\pm$ 39   \\
\end{tabular}
\end{ruledtabular}
\end{table*}

\begin{figure}[tbp]
\centering
\includegraphics[width=0.48\textwidth]{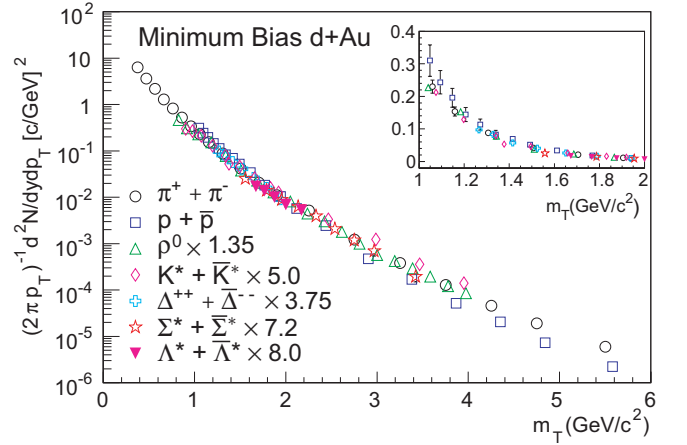}
\caption{(Color online) Proton and pion spectra \cite{ratios1}
plotted together with the rescaled minimum bias $d$+Au spectra of
$\rho^0$, $K^*$, $\Delta^{++}$, and $\Sigma^*$, where the
transverse mass scaling is observed. The inset plot is a zoom-in
of the region 1 $\leq m_T \leq$ 2 GeV/$c^2$ on a linear
scale.}\label{fig:scaling}
\end{figure}

The extracted $dN/dy$, $T$, and $n$ for the $\rho^0$ and the $K^*$
are listed in Table \ref{tab:rhokstardndy}. In the case of the
$\Delta^{++}$, $\Sigma^*$, and $\Lambda^*$, the corresponding
$dN/dy$ and $T$ are listed in Table
\ref{tab:deltasigmalambdadndy}. One contribution to the systematic
uncertainties quoted in Tables \ref{tab:rhokstardndy} and
\ref{tab:deltasigmalambdadndy} is due to the tracking efficiency
($\sim$8$\%$), which is common to all measurements.

In the case of the $\rho^0$, the normalization between the
$M_{\pi\pi}$ and the like-sign reference distributions is the
largest contribution to the systematic uncertainty to the yield
and the inverse slope ($T$) and it can be as large as
$\sim$20$\%$. If the $\rho^0$ invariant yield is obtained for the
case that the $\rho^0$ width is a free parameter in the hadronic
cocktail, the invariant yield increases by 22$\%$ from those shown
in Table \ref{tab:rhokstardndy}. In the other extreme, if the
invariant yield is obtained by assuming an exponential background,
the yields decrease by 45$\%$.

The contributions to the systematic uncertainty on the $K^*$ and
$\Delta^{++}$ yields and $T$ measured in $d$+Au collisions were
obtained by comparing different BW functions (relativistic and
non-relativistic), using different residual background functions
(first and second-order polynomial), different functions to fit
the spectra (exponential and power-law), and different slope
parameters in the BW$\times$PS function (140 MeV and 180 MeV). In
addition, the yields and $T$ were obtained separately for
$K^{*0}$, $\overline{K}^{*0}$, $K^{*+}$, $K^{*-}$, $\Delta^{++}$,
and $\overline{\Delta}^{--}$. The effect of opening the primary
vertex from 50 cm to 75 cm in the case of the yields obtained for
different centralities in $d$+Au collisions was also taken into
account. The systematic uncertainty on both yields and $T$ is
$\sim$ 20$\%$ for the $K^*$. In the case of the $\Delta^{++}$, the
systematic uncertainties are 12$\%$ and 17$\%$, respectively.

In minimum bias $p+p$ collisions, the main contributions to the
$\Delta^{++}$ yield and $T$ systematic uncertainty was estimated
from the invariant yields as a function of $p_T$ by increasing the
normalization between the $M_{p\pi}$ and the mixed-event reference
distributions until the fit to the $\Delta^{++}$ signal is not
reasonable. This procedure is then repeated by decreasing the
normalization. During this procedure, the width was fixed to 110
MeV/$c^2$ \cite{deltaformfactor}.

The number of partons (primarily gluons) in a nucleus grows very
rapidly at very high energies. If the occupation number of these
partons is large, they may saturate and form a novel state of
matter called a color glass condensate (CGC). This CGC has a bulk
scale which is the typical intrinsic transverse momentum of the
saturated gluons in the nucleus. The CGC can be probed in deep
inelastic scattering \cite{mcl,kovch}, in photo-production in
peripheral heavy-ion collisions \cite{gel}, in $p(d)$+A collisions
\cite{kov,dum} and in heavy-ion collisions
\cite{kharzeev,kovner,krasnitz}. Figure \ref{fig:scaling} shows that the
transverse mass ($m_T$) spectra of identified hadrons follow a generalized
scaling law in $d$+Au collisions between 1 $\leq
m_T \leq$ 2 GeV/$c^2$. Even though this scaling behavior is
motivated by the idea of a saturation of the gluon density, the
identified particle spectra measured in $p+p$ collisions at ISR
\cite{alper,alp,gat}, at Sp$\bar{\text{p}}$S \cite{jur1}, and at
RHIC \cite{starpp} energies have also been shown to follow a
generalized scaling law in transverse mass. More theoretical work
is needed in order to explain the similarities between $p+p$ and
$d$+Au collisions.

It is interesting to notice that for resonances in $d$+Au
collisions in the $p_T$ region measured, we do not observe the
shape difference of the $p_T$ spectra observed for mesons and
baryons in $p+p$ collisions at RHIC \cite{starpp}
for non-resonant particles in the interval $2 < p_T < 6 $ GeV/$c$
at the same beam energy. This baryon-meson effect observed in
$p+p$ collisions was argued to be a simple reflection of the
underlying dynamics of the collision in that meson production from
fragmentation requires only a (quark,anti-quark) pair while baryon
production requires a (di-quark,anti-di-quark) pair.

\subsection{Mean Transverse Momenta $\langle p_T \rangle$}

The averaged transverse momentum ($\langle p_T \rangle$) provides
information on the shape of the particle spectra. At a given mass,
the larger the $\langle p_T \rangle$, the harder the spectra are.
The resonance $\langle p_T \rangle$ were calculated from the fit
parameters depicted in Table \ref{tab:rhokstardndy} and Table
\ref{tab:deltasigmalambdadndy} and are listed in Table
\ref{tab:meanpt}.

\begin{table*}
\caption{\label{tab:meanpt}The $\rho^0$, $K^*$, $\Delta^{++}$,
$\Sigma^*$, and $\Lambda^*$ $\langle p_T \rangle$ in minimum bias
$d$+Au collisions and 0-20$\%$, 20-40$\%$, and 40-100$\%$ of the
total $d$+Au cross-section. The $\langle p_T \rangle$ measured in
minimum bias $p+p$ is also listed. The first error is
statistical; the second is systematic.}
\begin{ruledtabular}
\begin{tabular}{ccc}
Resonance & Centrality & $\langle p_T \rangle$ (GeV/$c$) \\
\hline
    $\rho^0$      & Minimum Bias $d$+Au &  0.808 $\pm$ 0.050 $\pm$ 0.086\\
                  & 0-20$\%$            &  0.815 $\pm$ 0.020 $\pm$ 0.083\\
                  & 20-40$\%$           &  0.805 $\pm$ 0.015 $\pm$ 0.082\\
                  & 40-100$\%$          &  0.764 $\pm$ 0.009 $\pm$ 0.081\\
                  & Minimum Bias $p+p$  &  0.616 $\pm$ 0.002 $\pm$ 0.062\\
  $K^*$           & Minimum Bias $d$+Au &  0.96 $\pm$ 0.02 $\pm$ 0.16\\
                  & 0-20$\%$            &  1.00 $\pm$ 0.07 $\pm$ 0.17\\
                  & 20-40$\%$           &  0.98 $\pm$ 0.05 $\pm$ 0.17\\
                  & 40-100$\%$          &  0.90 $\pm$ 0.03 $\pm$ 0.15\\
                  & Minimum Bias $p+p$  &  0.81  $\pm$ 0.02$\pm$ 0.14\\
   $\Delta^{++}$  & Minimum Bias $d$+Au &  0.89 $\pm$ 0.02 $\pm$ 0.14\\
                  & 0-20$\%$            &  0.98 $\pm$ 0.05 $\pm$ 0.16\\
                  & 20-40$\%$           &  0.92 $\pm$ 0.04 $\pm$ 0.18\\
                  & 40-100$\%$          &  0.90 $\pm$ 0.03 $\pm$ 0.14\\
                  & Minimum Bias $p+p$  &  0.63 $\pm$ 0.04 $\pm$ 0.13\\
    $\Sigma^*$    & Minimum Bias $d$+Au &  1.13  $\pm$ 0.03  $\pm$ 0.08\\
                  & 0-20$\%$            &  1.33 $\pm$ 0.06 $\pm$ 0.10\\
                  & 20-40$\%$           &  1.20 $\pm$ 0.07 $\pm$ 0.10\\
                  & 40-100$\%$          &  1.22 $\pm$ 0.07 $\pm$ 0.10\\
                  & Minimum Bias $p+p$  &  1.015 $\pm$ 0.015 $\pm$ 0.07\\
  $\Lambda^*$     & Minimum Bias $d$+Au &  1.17 $\pm$ 0.15 $\pm$ 0.12\\
                  & Minimum Bias $p+p$  &  1.08 $\pm$ 0.09 $\pm$ 0.05\\
\end{tabular}
\end{ruledtabular}
\end{table*}

\begin{figure}[tbp]
\centering
\includegraphics[width=0.45\textwidth]{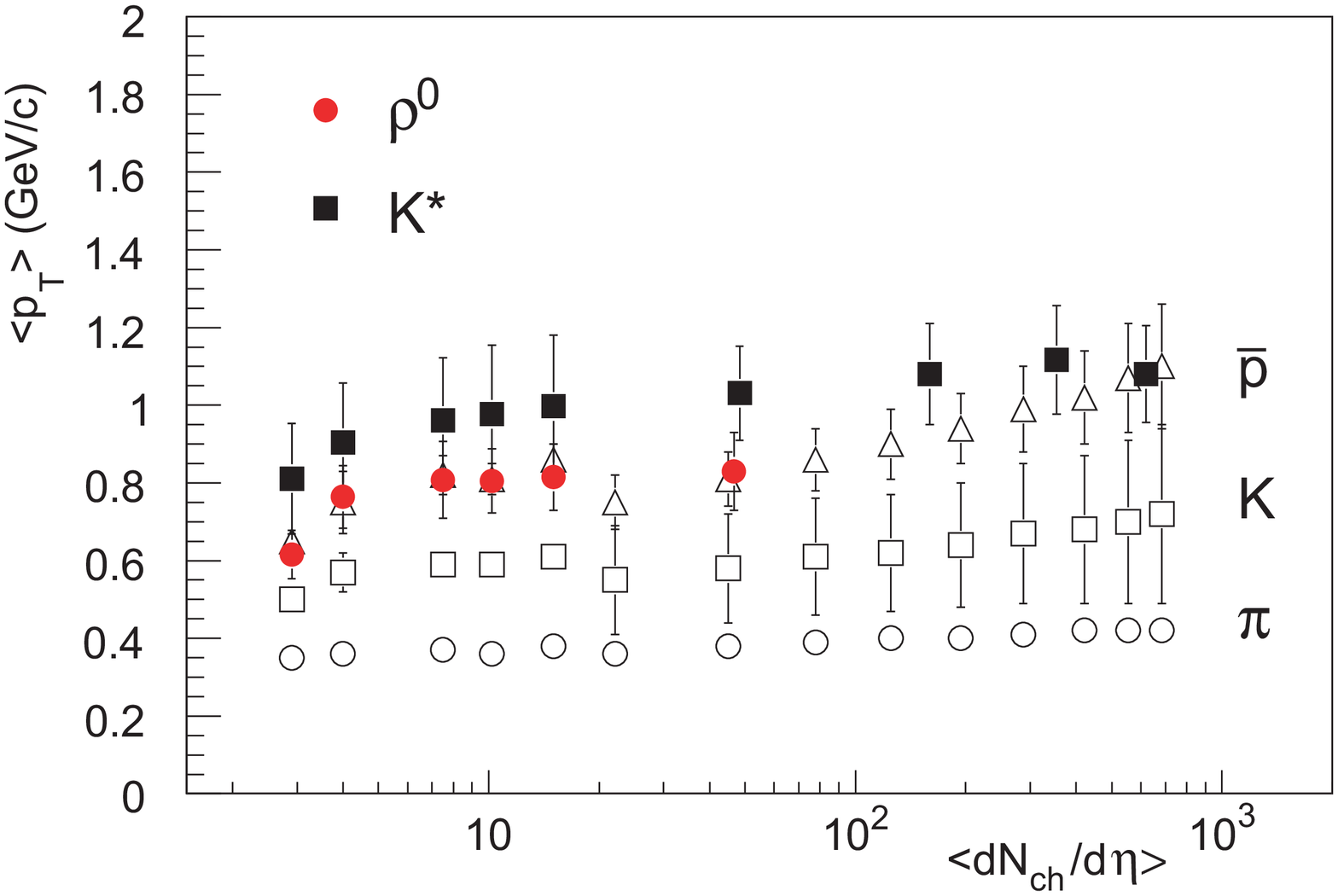}
\caption{(Color online) The $\rho^0$ and $K^*$ $\langle p_T
\rangle$ as a function of $\langle dN_{ch}/d\eta \rangle$ compared
to that of $\pi^-$, $K^-$, and $\overline{p}$ for minimum bias
$p+p$, minimum bias $d$+Au, and 0-20$\%$, 20-40$\%$, and
40-100$\%$ of the total $d$+Au cross-section \cite{meanpt}. The
errors shown are the quadratic sum of the statistical and
systematic uncertainties.}\label{fig:meanptrhokstar}
\end{figure}

\begin{figure}[tbp]
\centering
\includegraphics[width=0.45\textwidth]{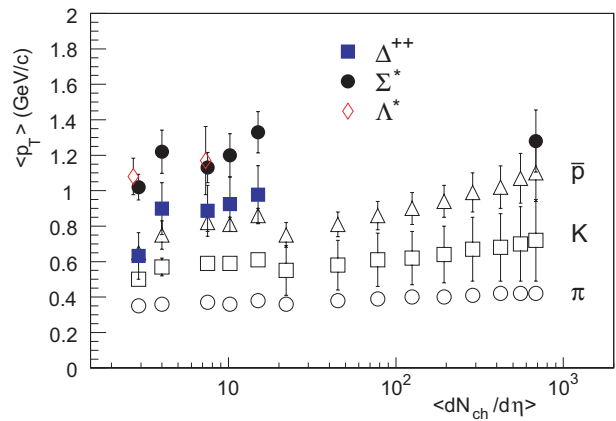}
\caption{(Color online) The $\Delta^{++}$, $\Sigma^*$, and
$\Lambda^*$ $\langle p_T \rangle$ as a function of $\langle
dN_{ch}/d\eta \rangle$ compared to that of $\pi^-$, $K^-$, and
$\overline{p}$ for minimum bias $p+p$, minimum bias $d$+Au, and
0-20$\%$, 20-40$\%$, and 40-100$\%$ of the total $d$+Au
cross-section \cite{meanpt}. The errors shown are the quadratic
sum of the statistical and systematic
uncertainties.}\label{fig:meanptdeltasigma}
\end{figure}

The $\rho^0$, $K^*$, $\Delta^{++}$, $\Lambda^*$, and $\Sigma^*$
$\langle p_T \rangle$ as a function of $dN_{ch}/d\eta$ compared to
that of $\pi^-$, $K^-$, and $\overline{p}$ for minimum bias $d$+Au
\cite{meanpt} are depicted in Figs. \ref{fig:meanptrhokstar} and
\ref{fig:meanptdeltasigma}. While the $\langle p_T \rangle$ of
these hadrons are independent of centrality, as expected, the
$\langle p_T \rangle$ is strongly dependent on the mass of the
particle.

\begin{figure}[tbp]
\centering
\includegraphics[width=0.45\textwidth]{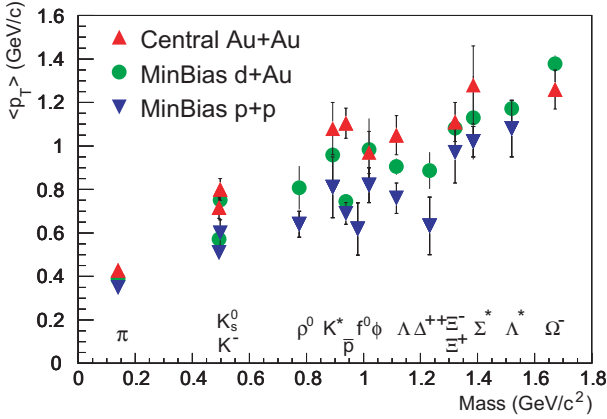}
\caption{(Color online) The $\langle p_T \rangle$ of various
particles for different systems, minimum bias $p+p$, $d$+Au, and
central Au+Au collisions. The errors shown are the quadratic sum
of the statistical and systematic
uncertainties.}\label{fig:daumeanpt}
\end{figure}

We can compare the spectra shape among particles for different
systems by comparing their $\langle p_T \rangle$. Figure
\ref{fig:daumeanpt} shows the $\langle p_T \rangle$ of various
particles for different systems, minimum bias $p+p$, $d+Au$, and
central Au+Au collisions. Even though there is a strong mass
dependence, the $\langle p_T \rangle$ of these particles do not
appear to strongly depend on the collision system, with the
exception of the $\bar{p}$. However, the $\langle p_T \rangle$ of
particles measured in $d$+Au collisions lie between the $\langle
p_T \rangle$ measured in $p+p$ and Au+Au collisions, indicating a
hardening of the spectra from $p+p$ through $d$+Au to Au+Au
collisions.

The main contributions to the systematic uncertainties quoted in
Table \ref{tab:meanpt} are due to tracking efficiency
($\sim$8$\%$) and different fit functions used to fit the $p_{T}$
spectra. In the case of the $\rho^0$, in addition there was the
normalization between the $\pi^+\pi^-$ invariant mass distribution
and the like-sign reference distributions ($\sim$5$\%$).

\subsection{Particle Ratios}

\begin{figure}[tbp]
\centering
\includegraphics[width=0.45\textwidth]{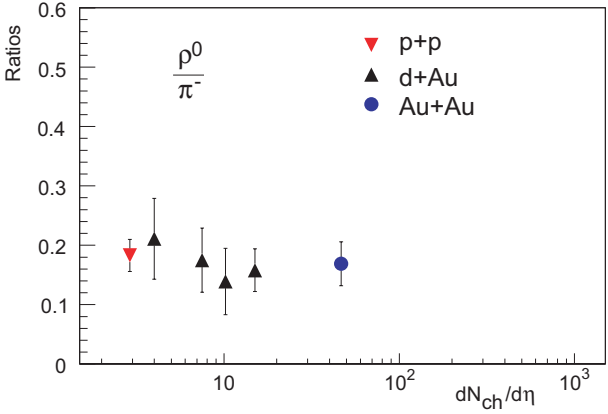}
\caption{(Color online) The $\rho^0/\pi^-$ ratios in $p+p$, various centralities in
$d$+Au, and in peripheral $Au+Au$ collisions as a function of $dN_{ch}/d\eta$.
The errors shown are the
quadratic sum of the statistical and systematic
uncertainties.}\label{fig:ratiosrho}
\end{figure}

\begin{figure}[tbp]
\centering
\includegraphics[width=0.45\textwidth]{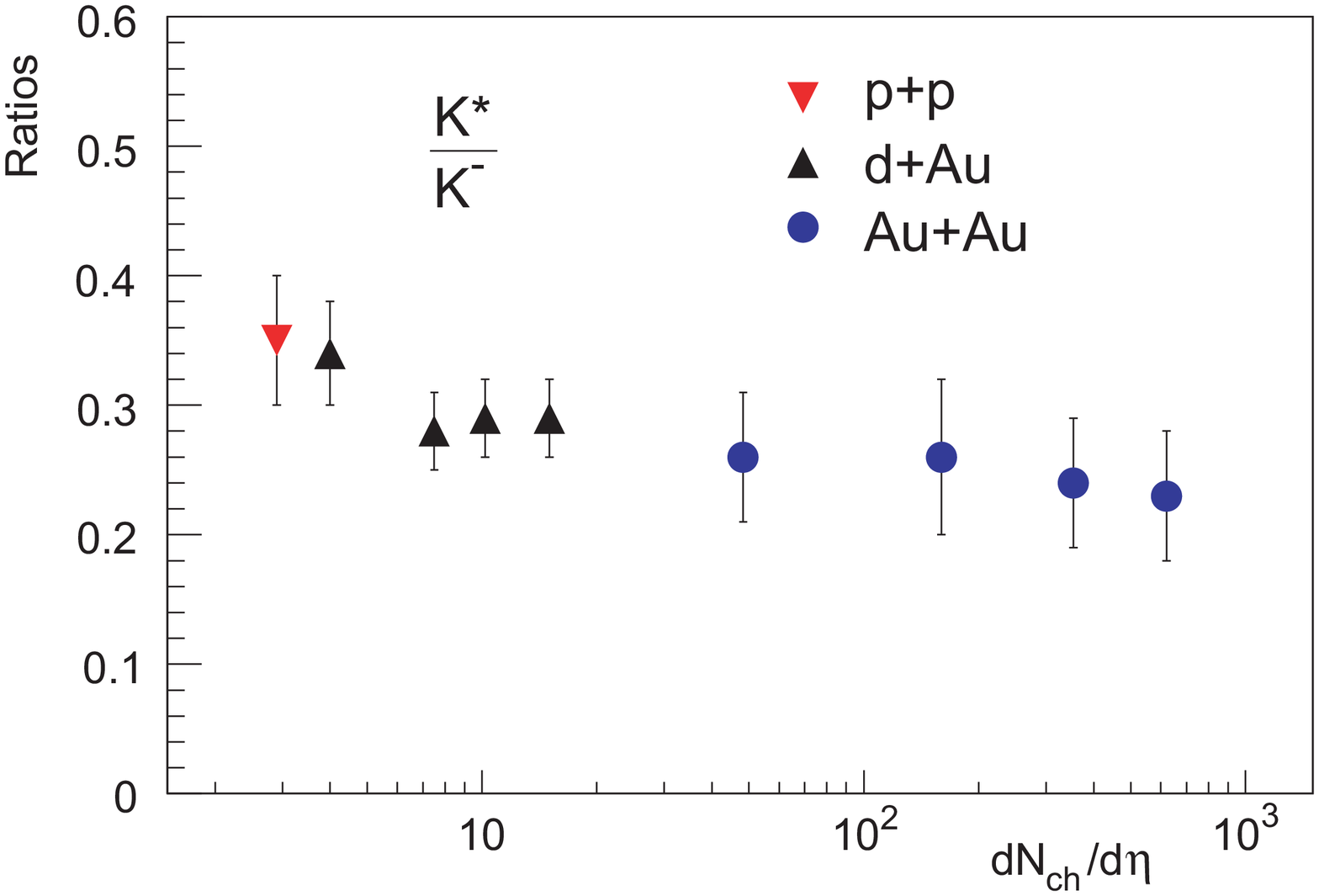}
\caption{(Color online) The $K^*/K^-$ ratios in $p+p$ and various centralities in $d$+Au
and Au+Au collisions as a function of $dN_{ch}/d\eta$. The errors shown are the
quadratic sum of the statistical and systematic
uncertainties.}\label{fig:ratioskstar}
\end{figure}

\begin{figure}[tbp]
\centering
\includegraphics[width=0.45\textwidth]{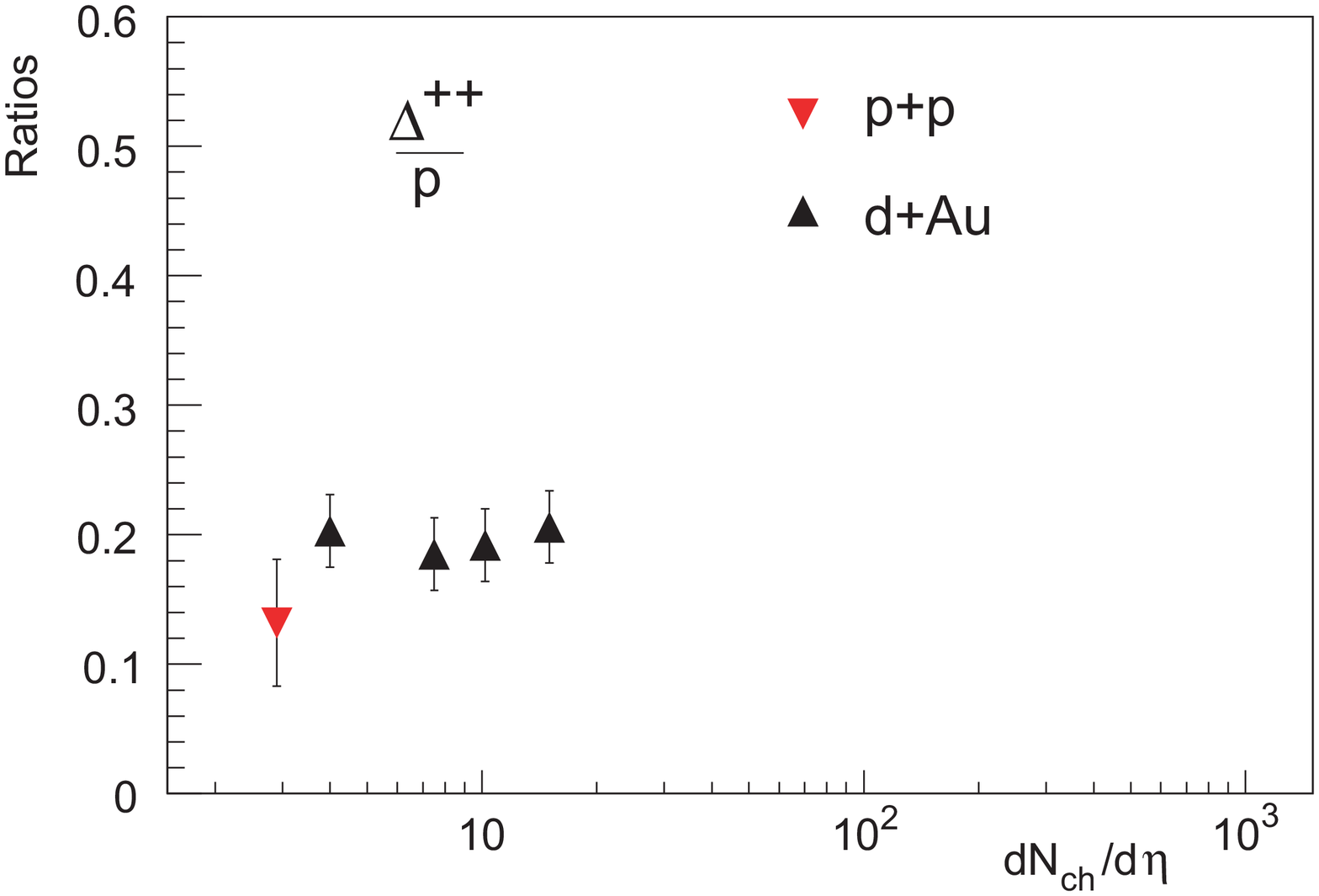}
\caption{(Color online) The $\Delta^{++}/p$ ratios in $p+p$ and various centralities in
$d$+Au collisions as a function of $dN_{ch}/d\eta$. The errors shown are the
quadratic sum of the statistical and systematic
uncertainties.}\label{fig:ratiosdelta}
\end{figure}

\begin{figure}[tbp]
\centering
\includegraphics[width=0.45\textwidth]{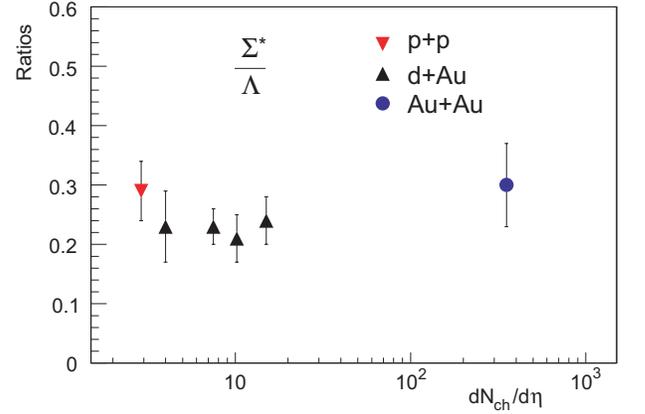}
\caption{(Color online) The $\Sigma^{*}/\Lambda$ ratios in $p+p$, various
centralities in $d$+Au, and in central Au+Au collisions as a function
of $dN_{ch}/d\eta$. The errors shown are
the quadratic sum of the statistical and systematic
uncertainties.}\label{fig:ratiossigma}
\end{figure}

\begin{figure}[tbp]
\centering
\includegraphics[width=0.45\textwidth]{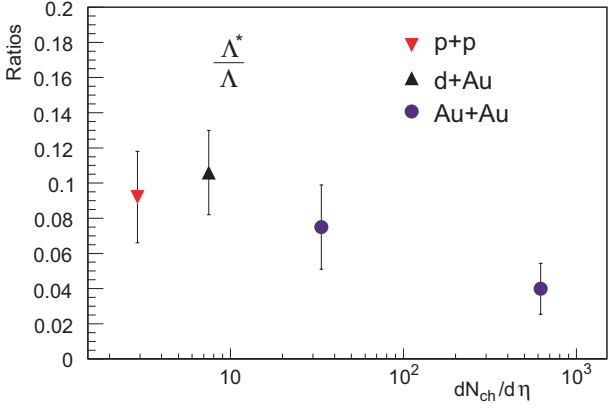}
\caption{(Color online) The $\Lambda^{*}/\Lambda$ ratios in $p+p$,
minimum bias $d$+Au, and two different centralities in Au+Au collisions
as a function of $dN_{ch}/d\eta$. The errors
shown are the quadratic sum of the statistical and systematic
uncertainties.}\label{fig:ratioslambda}
\end{figure}

It has been previously shown that the ratio of yields of resonances
to the yields of stable particles can effectively probe the
dynamics of relativistic heavy-ion collisions
\cite{kstar130,rho,kstar200,baryon}. The ratios of yields of
resonances to stable particles with similar quark content but
different spin and masses are given in Table \ref{tab:ratios}. The
values of $\pi, K,$ and $p$ were taken from \cite{ratios1,ratios}.
Figures \ref{fig:ratiosrho}, \ref{fig:ratioskstar},
\ref{fig:ratiosdelta}, \ref{fig:ratiossigma} and
\ref{fig:ratioslambda} show the ratios of resonances to their
corresponding stable particles as a function of the charged
particle multiplicity ($N_{ch}$) in $p+p$, $d$+Au and Au+Au
collisions. We observe that the $\rho^0/\pi^-$, $\Delta^{++}/p$,
and $\Sigma^{*}/\Lambda$ ratios are independent of multiplicity
while the $K^*/K^-$ and $\Lambda^{*}/\Lambda$ ratios seem to decrease. The
resonance abundance could be affected by mass shifts due to phase
space ($\exp(-m/T)$) in different collision systems.

\begin{figure}[tbp]
\centering
\includegraphics[width=0.45\textwidth]{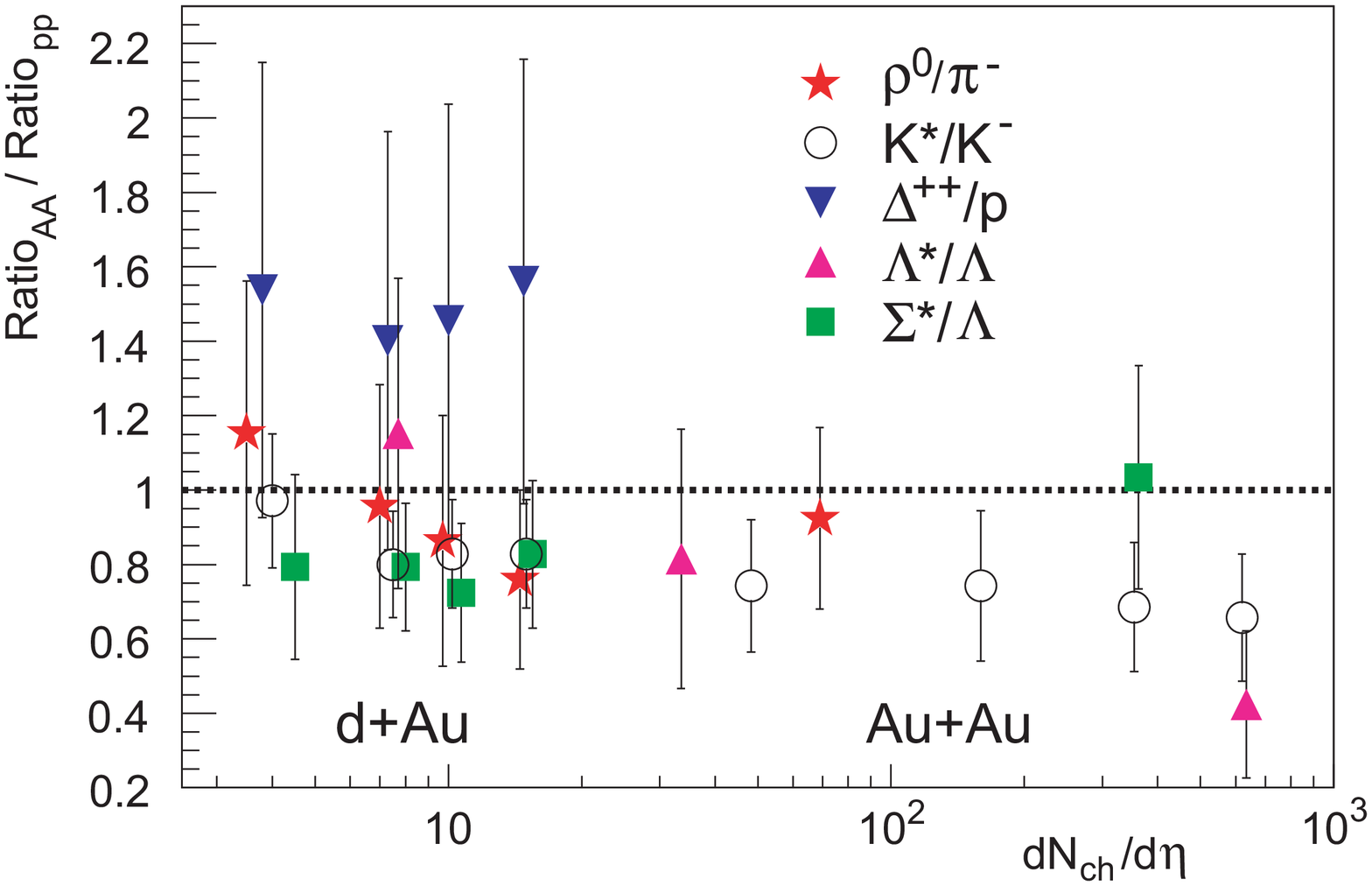}
\caption{(Color online) The resonance ratios normalized by their
ratio measured in $p+p$ collisions at the same beam energy as a
function of $dN_{ch}/d\eta$. The errors shown are the quadratic
sum of the statistical and systematic
uncertainties.}\label{fig:ratios}
\end{figure}

The resonance ratios normalized by their value measured in $p+p$
collisions at the same $\sqrt{s}$ are plotted in Fig.
\ref{fig:ratios}. The decrease of the resonance ratios of
$K^*/K^-$ and $\Lambda^{*}/\Lambda$ from $p+p$ to Au+Au collisions
has been explained by an extended lifetime of the hadronic phase
where the re-scattering of the decay particles dominates over
resonance regeneration
\cite{kstar130,kstar200,baryon,urqmd1,urqmd2,mar02}. As the
$K^*/K^-$ and the $\Lambda^{*}/\Lambda$ ratios are similar for
$p+p$ and $d$+Au collisions, this would suggest the absence of an
extended hadronic medium in $d$+Au collisions. The $\rho^0/\pi^-$,
$\Delta^{++}/p$ and $\Sigma^{*}/\Lambda$ ratios in $d$+Au
collisions are in agreement with their ratios measured in $p+p$
collisions. These resonance ratios do not show any suppression
from $p+p$ to Au+Au collisions either, hence they are not
sensitive to the lifetime of the hadronic medium, presumably due
to their large regeneration cross-section.

The $\rho^0/\pi^-$ ratio is independent of centrality up to the
40-80$\%$ of the inelastic hadronic Au+Au cross-section and it is
of the same order as the corresponding $p+p$ measurement. In $p+p$
collisions, it has been proposed that the mass shift is due to
$\pi\pi$ re-scattering, even in the absence of a medium
\cite{fac}. If this is the case, $\pi^+\pi^-$ re-scattering might
regenerate the $\rho^0$. In addition, one of the decay daughters
might also re-scatter with other hadrons preventing the $\rho^0$
to be measured. Therefore, these two processes compete with (and
balance) each other.

\begin{table*} \caption{\label{tab:ratios}The $\rho^0/\pi^-$,
$K^*/K^-$, $\Delta^{++}/p$, $\Sigma^*/\Lambda$, and
$\Lambda^*/\Lambda$ ratios in minimum bias $p+p$
\cite{rho,kstar200,baryon}, $d$+Au, and 0-20$\%$, 20-40$\%$, and
40-100$\%$ of the total $d$+Au cross-section. The first error is
statistical; the second is systematic.}
\begin{ruledtabular}
\begin{tabular}{cccccc}
 Centrality & $\rho^0/\pi^-$ & $K^*/K^-$ & $\Delta^{++}/p$ & $\Sigma^*/\Lambda$
 & $\Lambda^*/\Lambda$ \\ \hline

 Min. Bias $d$+Au     & 0.175 $\pm$ 0.004 $\pm$ 0.054 & 0.28 $\pm$ 0.01 $\pm$ 0.03 &
 0.185 $\pm$ 0.005 $\pm$ 0.028 & 0.23 $\pm$ 0.03 & 0.106 $\pm$ 0.024 \\

 0-20$\%$                & 0.139 $\pm$ 0.014 $\pm$ 0.036 & 0.29 $\pm$ 0.01 $\pm$ 0.03 &
 0.206 $\pm$ 0.006 $\pm$ 0.028 & 0.24 $\pm$ 0.04 &  \\

 20-40$\%$               & 0.158 $\pm$ 0.011 $\pm$ 0.056 & 0.29 $\pm$ 0.01 $\pm$ 0.03 &
 0.192 $\pm$ 0.005 $\pm$ 0.028 & 0.21 $\pm$ 0.04 &  \\

 40-100$\%$              & 0.211 $\pm$ 0.005 $\pm$ 0.068 & 0.34 $\pm$ 0.01 $\pm$ 0.04 &
 0.203 $\pm$ 0.006 $\pm$ 0.028 & 0.23 $\pm$ 0.06 &  \\

 Min. Bias $p+p$      & 0.183 $\pm$ 0.001 $\pm$ 0.027 & 0.35 $\pm$ 0.01 $\pm$ 0.05 &
 0.132 $\pm$ 0.002 $\pm$ 0.049 & 0.029 $\pm$ 0.047 &  0.092 $\pm$ 0.026\\

\end{tabular}
\end{ruledtabular}
\end{table*}

\subsection{Nuclear Modification Factor}

The nuclear modification factor ($R_{dAu}$) is defined as
\begin{eqnarray}
R_{dAu}(p_T) = \frac{d^2N_{dAu}/dydp_T}{\langle N_{bin}
\rangle/\sigma_{pp}^{inel} \times d^2\sigma_{pp}/dydp_T},
\end{eqnarray}
where $d^2N_{dAu}/dydp_T$ is the yield of the produced
particles in minimum bias $d$+Au collisions,
$d^2\sigma_{pp}/dydp_T$ is the inclusive cross-section in $p+p$ collisions, $\langle
N_{bin} \rangle$ is the average number of binary nucleon-nucleon
($NN$) collisions per event, and $\langle N_{bin}
\rangle/\sigma_{pp}^{inel}$ is the nuclear overlap $T_A(b)$
\cite{cent,rdau,rdau1}. The value of $\sigma_{pp}^{inel}$ is 42
mb.

The enhancement observed in $R_{dAu}$ for high $p_T$ and
mid-rapidity, known as the Cronin effect \cite{cronin}, is
generally attributed to the influence of multiple parton
scattering through matter prior to the hard scattering that
produces the observed high-$p_T$ hadron \cite{pscatt}. Therefore,
the nuclear modification factor ($R_{dAu}$) can be used to study
the effects of matter on particle production.

\begin{figure}[tbp]
\centering
\includegraphics[width=0.45\textwidth]{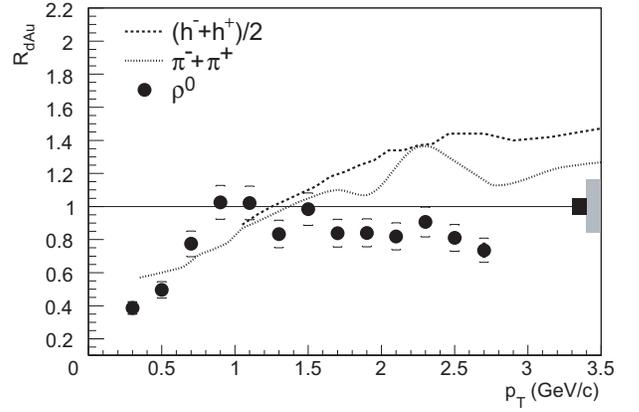}
\caption{The $\rho^0$ $R_{dAu}$ compared to the charged hadrons
$R_{dAu}$. The shaded box is the error on the overall
normalization and the black box is the error on
$N_{bin}$.}\label{fig:raarho}
\end{figure}

\begin{figure}[tbp]
\centering
\includegraphics[width=0.45\textwidth]{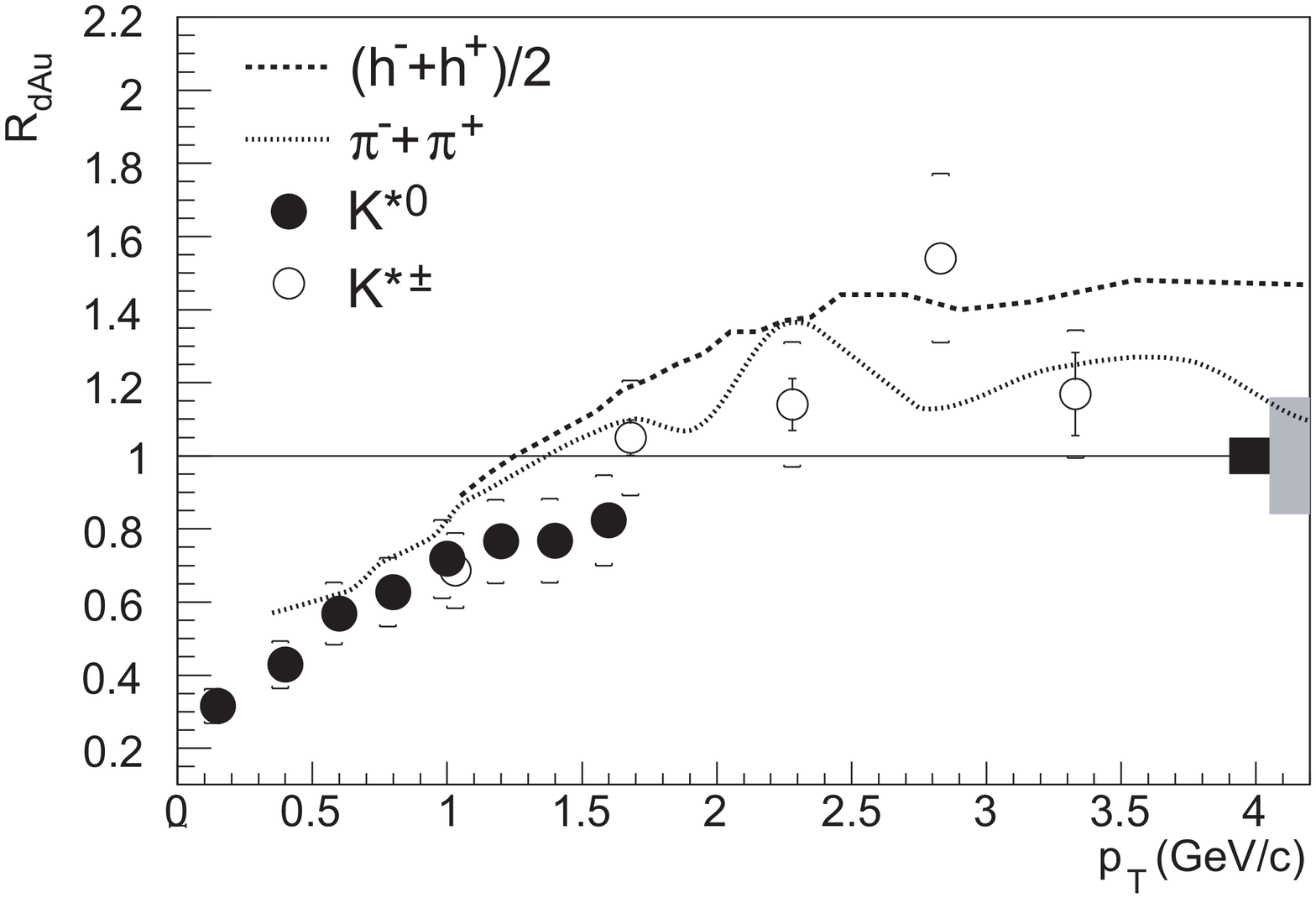}
\caption{The $K^*$ $R_{dAu}$ compared to the charged hadrons
$R_{dAu}$. The shaded box is the error on the overall
normalization and the black box is the error on
$N_{bin}$.}\label{fig:raakstar}
\end{figure}

\begin{figure}[tbp]
\centering
\includegraphics[width=0.45\textwidth]{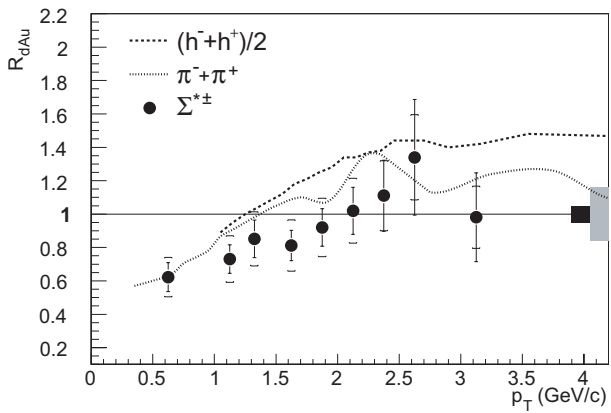}
\caption{The $\Sigma^*$ $R_{dAu}$ compared to the charged hadrons
$R_{dAu}$. The shaded box is the error on the overall
normalization and the black box is the error on
$N_{bin}$.}\label{fig:raasigma}
\end{figure}

The $R_{dAu}$ for $\rho^0$, $K^*$, and $\Sigma^*$ are shown in
Figs. \ref{fig:raarho}, \ref{fig:raakstar}, and \ref{fig:raasigma}
together with the $R_{dAu}$ of charged hadrons and charged pions. The
$K^*$ and $\Sigma^*$ $R_{dAu}$ are lower than unity at low $p_T$
and consistent with the $R_{dAu}$ of charged hadrons and charged
pions. The $R_{dAu}$ of the $\rho^0$, $K^*$, and $\Sigma^*$  scale
with $N_{bin}$ for $p_T >$ 1.2 GeV/$c$ taking into account the
uncertainties in the normalization. We also observe that the
$\rho^0$ $R_{dAu}$ for $p_T >$ 1.5 GeV/$c$ is suppressed compared
to the charged hadrons and charged pions $R_{dAu}$. The
$\Delta^{++}$ $R_{dAu}$ is not shown due to the small $p_T$ range
covered and the large uncertainties in the measurement.

More information may be obtained from the $R_{dAu}$ measurement if it is
extended to higher $p_T$. In STAR this will be possible with the installation of the barrel Time-of-Flight (TOF) detector. The TOF will provide essential particle identification by, for instance, increasing the percentage of kaon and protons for which particle identification is possible to more than 95$\%$ of all those produced within the acceptance of the TOF barrel ($|\eta| \leq$ 1.0). The improvement
in particle identification will allow a decrease in the signal to background ratios for the
resonance measurements.

\section{Summary}

Measurements of $\rho(770)^0$, $K^*$(892), $\Delta$(1232)$^{++}$,
$\Sigma(1385)$, and $\Lambda(1520)$ in $\sqrt{s_{_{NN}}}$ = 200
GeV $d$+Au collisions reconstructed via their hadronic decay
channels using the STAR detector are presented.

The masses of the $\rho^0$, $K^*$, $\Delta^{++}$, $\Sigma(1385)$,
and $\Lambda$(1520) are measured for minimum bias and three
different centralities in $d$+Au collisions at $\sqrt{s_{NN}}$ =
200 GeV. We observe a $\rho^0$ mass shift of $\sim$50 MeV/$c^2$ at
low $p_T$. In addition, the $\rho^0$ mass measured in 0-20$\%$ of
the total $d$+Au cross-section is slightly lower than the mass
measured in the most peripheral centrality class. We also observe
that the $\rho^0$ mass measured in minimum bias $d$+Au and high
multiplicity $p+p$ interactions are comparable. The $K^{*0}$ and
$\Sigma^*$ masses at low $p_T$ ($p_T < $1.1 GeV/$c$) are smaller
than previously measured values \cite{pdg} by up to $\sim$10
MeV/$c^2$. A similar mass shift for the $K^{*0}$ is observed in
minimum bias $p+p$ collisions at $\sqrt{s_{NN}}$ = 200 GeV
\cite{kstar200}. The $K^{*\pm}$ mass and the $K^*$ width are in
agreement with previously reported values
 within errors \cite{pdg}. The $\Delta^{++}$ mass is
shifted by $\sim$40 MeV/$c^2$ in minimum bias $p+p$ and $\sim$50
MeV/$c^2$ in minimum bias $d$+Au collisions. In contrast to the
$\rho^0$, no $p_T$ dependence is observed. Similar mass and/or
width modifications with respect to
those observed in e+e- collisions are observed for these resonances in $p+p$ and Au+Au collisions. The possible
explanations for the apparent modification of the $\rho^0$ meson
properties are interference between various $\pi^+\pi^-$
scattering channels, phase space distortions due to the
re-scattering of pions forming $\rho^0$, and Bose-Einstein
correlations between $\rho^0$ decay daughters and pions in the
surrounding matter \cite{rho}. All these explanations require an
interaction, which implies a medium such as the one formed in A+A
collisions. However, the $\rho^0$ mass shift measured in $p+p$
collisions \cite{rho} can be described without a medium
\cite{fac}. The $\Delta^{++}$ width and the $\Lambda^*$ mass and
width measured are in agreement with previous measurements
\cite{pdg}.

The transverse mass spectra follows a generalized scaling law between 1 and 2 GeV/$c^2$. However, in Au+Au collisions at RHIC at $\sqrt{s_{NN}}$ =
200 and 62.4 GeV a generalized scaling law is not observed \cite{star62200}, possibly due to additional physics effects such as flow, coalescence and energy loss. Even though the scaling behavior in d+Au collisions is
motivated by the idea of a saturation of the gluon density, the
identified particle spectra measured in $p+p$ have also been shown to follow a
generalized scaling law in transverse mass. More theoretical work
is needed in order to explain the similarities between $p+p$ and
$d$+Au collisions. The resonances in $d$+Au
collisions in the measured $p_T$ region do not show the
shape difference of the $p_T$ spectra observed for mesons and
baryons in $p+p$ collisions at RHIC
for non-resonant particles in the interval $2 < p_T < 6 $ GeV/$c$
at the same beam energy. This baryon-meson effect observed in
$p+p$ collisions was argued to be a simple reflection of the
underlying dynamics of the collision. In order to have further insight in this matter, the spectra of resonances should be increased to higher momentum, which will be accomplished in STAR with the
installation of the barrel Time-of-Flight that will provide extended particle identification.

The $\rho^0$, $K^*$, $\Delta^{++}$, $\Sigma^*$, and $\Lambda^*$
$\langle p_T \rangle$ are found to be centrality independent.
Compared to the $\langle p_T \rangle$ of pions, kaons, and
anti-protons, we measure that the $\langle p_T \rangle$ of these
resonances are approximately the same as or even higher than the
proton $\langle p_T \rangle$. The resonances $\langle p_T \rangle$
as a function of centrality follow a mass ordering.

The $\rho^0/\pi^-$, $K^*/K^-$, $\Delta^{++}/p$,
$\Sigma^*/\Lambda$, and $\Lambda^*/\Lambda$ ratios are measured
and we find that the $\rho^0/\pi^-$ ratio is independent of
centrality up to the 40-80$\%$ of the inelastic hadronic Au+Au
cross-section and it is of the same order of the corresponding
$p+p$ measurement. If we speculate there is particle
re-scattering even without the presence of a medium for
short-lived resonances. We can interpret these results in terms of
the re-scattering/regeneration scenario and conclude that in both
cases the regeneration is the dominant process. We observe that
the $K^*/K^-$ ratio is the same for $p+p$ and the most peripheral
centrality class in $d$+Au collisions. Then, it slightly decreases
to peripheral Au+Au collisions to a suppression in central Au+Au
collisions, showing that the re-scattering is the dominant
process. The $\Sigma^{*}/\Lambda$ and the $\Lambda^{*}/\Lambda$
ratios measured in $d$+Au collisions are the same as those
measured in $p+p$ collisions within errors, as expected since they
are not as short-lived as the $\rho^0$, $K^*$ or $\Delta^{++}$.

The $R_{dAu}$ of the $\rho^0$, $K^*$, and $\Sigma^*$  scale with
$N_{bin}$ for $p_T >$ 1.2 GeV/$c$ taking into account the
uncertainties in the normalization. We also observed that the
$\rho^0$ $R_{dAu}$ for $p_T >$ 1.5 GeV/$c$ is suppressed compared
to the charged hadrons and charged pions $R_{dAu}$. More information may be obtained from the $R_{dAu}$ measurement if it is
extended to higher $p_T$ which will be accomplished in STAR with the installation of the TOF.

The measurement of these resonances in $d$+Au collisions will
provide reference for future measurements in A+A collisions.

\section{Acknowledgments}

We thank the RHIC Operations Group and RCF at BNL, and the NERSC
Center at LBNL for their support. This work was supported in part
by the HENP Divisions of the Office of Science of the U.S. DOE;
the U.S. NSF; the BMBF of Germany; IN2P3, RA, RPL, and EMN of
France; EPSRC of the United Kingdom; FAPESP of Brazil; the Russian
Ministry of Science and Technology; the Ministry of Education and
the NNSFC of China; IRP and GA of the Czech Republic, FOM of the
Netherlands, DAE, DST, and CSIR of the Government of India; Swiss
NSF; the Polish State Committee for Scientific Research; STAA of
Slovakia, and the Korea Sci. $\&$ Eng. Foundation.

\end{document}